\documentclass[aps,pra,showpacs,twocolumn,superscriptaddress]{revtex4-1}
\usepackage{bm,color,amsmath,amssymb,mathrsfs,latexsym,graphicx,psfrag,bbm}
\usepackage[T1]{fontenc}

\newcommand{\PlowE}[0]{\mathcal{P}_{\mbox{\tiny lowE}}}

\newcommand{\ket}[1]{|#1\rangle}
\newcommand{\ev}[1]{\langle #1 \rangle}

\newcommand{\bk}[0]{\mathbf k}
\newcommand{\bp}[0]{\mathbf p}
\newcommand{\br}[0]{\mathbf r}
\newcommand{\bra}[1]{\langle #1|}
\newcommand{\braket}[2]{\langle #1|#2\rangle}

\newcommand{\spl}[1]{\begin{align}\begin{split} #1 \end{split} \end{align}}

\newcommand{\bcut}[0]{\alpha_\text{max}}

\begin{document}
\title{Effective multibody-induced tunneling and interactions in the Bose-Hubbard model \\ of the lowest dressed band of an optical lattice}
\author{Ulf Bissbort}
\affiliation{Institut f\"ur Theoretische Physik, Johann Wolfgang Goethe-Universit\"at, 60438 Frankfurt/Main, Germany}
\author{Frank Deuretzbacher}
\affiliation{Institut f\"ur Theoretische Physik, Johann Wolfgang Goethe-Universit\"at, 60438 Frankfurt/Main, Germany}
\author{Walter Hofstetter}
\affiliation{Institut f\"ur Theoretische Physik, Johann Wolfgang Goethe-Universit\"at, 60438 Frankfurt/Main, Germany}

\begin{abstract}
We construct the effective lowest-band Bose-Hubbard model incorporating interaction-induced on-site correlations. The model is based on ladder operators for local correlated states, which deviate from the usual Wannier creation and annihilation, allowing for a systematic construction of the most appropriate single-band low-energy description in the form of the extended Bose-Hubbard model. A formulation of this model in terms of ladder operators not only naturally contains the previously found effective multibody interactions, but also contains \emph{multibody-induced} single-particle tunneling, pair tunneling and nearest-neighbor interaction processes of higher orders. An alternative description of the same model can be formulated in terms of occupation-dependent Bose-Hubbard parameters. These multi-particle effects can be enhanced using Feshbach resonances, leading to corrections which are well within experimental reach and of significance to the phase diagram of ultracold bosonic atoms in an optical lattice. We analyze the energy reduction mechanism of interacting atoms on a local lattice site and show that this cannot be explained only by a spatial broadening of Wannier orbitals on a single-particle level, which neglects correlations.
\end{abstract}

\maketitle
\section{Introduction}
Ultracold atoms in optical lattices are an ideal testing ground for models in solid-state physics due to the large degree of control over external and internal parameters of these many-body systems \cite{Bloch2008}. On the one hand, these systems are very promising as analog quantum simulators for gaining a further understanding of complicated solid state systems \cite{Jaksch98,Feynman82,Hofstetter02,Esslinger10}, whereas, on the other hand, completely new models (e.g., with further internal degrees of freedom, different quantum statistics, etc.) can be realized in a very clean and controlled fashion. Specifically, a large focus has been on ultracold bosonic atoms in optical lattices, which are well described by the Bose-Hubbard model \cite{Jaksch98}. The first milestone was the experimental observation of the superfluid-Mott insulator transition \cite{Greiner02}. With the ever increasing precision in recent experiments \cite{Will10, Will11, Trotzky10}, as well as the development of new probing techniques and remarkable technical advances \cite{Ospelkaus06, Best09, Mark11, Bakr11}, it has become possible to observe effects beyond the standard Hubbard model. Specifically, a density dependence of the interaction parameter $U$ has been observed by using quantum phase revival spectroscopy \cite{Will10, Will11}, which has been predicted and described using effective many-body interactions. A recent experiment using multiband spectroscopy to investigate the effect of bosons in a Bose-Fermi mixture found a significant reduction of the fermionic tunneling energy $J$ \cite{Heinze11}.

While the Fock space spanned by the Fock states generated by the full multiband single-particle Wannier orbitals is a perfectly valid basis for the interacting many-body system, where, by construction, the parameters are density-independent \cite{Larson2009}, it is customary to work in an effective single-band basis. However, such a description requires a density dependence of the parameters, or, alternatively, the introduction of effective higher-order terms, as will be shown.

It has been proposed \cite{Luehmann08, Lutchyn09, Mering11} that the density dependence of the bosonic tunneling parameter induced by the Bose-Fermi interaction can explain the shift in the bosonic superfluid-Mott insulator transition observed in Bose-Fermi mixtures \cite{Ospelkaus06, Best09}. This topic is still under debate, with an alternative cause suggested to be the heating of the system as the lattice is ramped up \cite{Cramer08, Cramer11, Snoek11}. Furthermore, several new phases have recently been predicted for the effective single-band density-dependent Bose-Hubbard model \cite{Hazzard10, Dutta11}. An effective density-dependent change of the Hubbard parameters has been calculated using a mean-field decoupling of the densities in Bose-Fermi mixtures \cite{Luehmann08,Lutchyn09} and also beyond this approximation \cite{Mering11}, where two-particle hopping amplitudes and further relevant Bose-Fermi Hubbard parameters were calculated within the full multi-orbital picture. In a single species bosonic lattice gas, the density-dependence of $J$ and the on-site interactions $U$ were calculated by minimizing the energy with respect to the real-space Wannier orbitals within a mean-field approach \cite{Li2006}. These, as well as nearest-neighbor interactions, were also determined within a Gaussian approximation for the Wannier functions \cite{Hazzard10}. The density-dependence of the single-particle tunneling amplitude $J$ and the interaction parameter $U$, as well as the effect on the phase diagram were considered in \cite{Dutta11,Luehmann11}. In \cite{Luehmann11}, a fully correlated, multi-orbital calculation was performed in the Wannier basis to quantitatively determine the density-dependence of $J$. Using a set of orthogonal variational orbitals and minimizing the energy with respect to their real-space shape and occupation number, the superfluid-Mott insulator transition in an interacting one-dimensional (1D) gas in an optical lattice was determined in \cite{Alon2005}. In the noninteracting Wannier or Bloch basis, this multi-orbital mean-field approach thus intrinsically contains higher band contributions.

In this work, we rigorously derive and define the effective lowest-band representation used in these previous works, where the localized many-body low-energy states are \emph{dressed} with contributions from higher bands, analogous to the dressed state basis in quantum optics. We define new ladder operators connecting only states within this dressed low-energy manifold, which exactly fulfill bosonic commutation relations. For finite interaction strength $|g|>0$, these do not coincide with the usual single-particle Wannier creation and annihilation operators and we give the exact prescription for transforming operators between the multi-orbital Wannier and the dressed single-band basis in the low-energy description. This transformation is also vital to translate any operator into the new basis, which is usually given in the real-space, Bloch or Wannier representation, e.g., observables, additional terms in the Hamiltonian, or perturbations. On a local level, our transformation recovers the effective multibody interactions found in \cite{Johnson09} in the limit of strong lattice depths $s$, where a Gaussian approximation for the Wannier functions applies. Furthermore, our basis transformation procedure allows for a systematic treatment of all nonlocal terms. These have been addressed in the context of Bose-Fermi mixtures in \cite{Mering11} and identify the counterparts of local multibody interactions: multi-particle-induced tunneling and correlated-pair tunneling terms arising from the usual bosonic interacting lattice Hamiltonian.

This paper is organized as follows: in Sec.~(\ref{SEC:2_alt_description}) we juxtapose the multi-orbital Wannier and the effective single-band descriptions and introduce the basis states of the latter. In Secs.~(\ref{SEC:low_E_subspace}) and (\ref{SEC:basis_transformation}) we define the low-energy subspace and the new effective bosonic ladder operators, from which the transformation properties are derived. Subsequently, they are applied in the systematic derivation of additional terms to the standard Bose-Hubbard model and are shown to give rise to $n$-particle-induced single-particle and correlated two-particle tunneling in  in Sec.~(\ref{SEC:single_particle_tunneling}) and (\ref{SEC:two_particle_tunneling}) respectively, as well as multibody nearest-neighbor interactions in Sec.~(\ref{SEC:nn_interactions}). Finally, we investigate the main energy reduction mechanism in Sec.~(\ref{SEC:higher_order_correlations}), showing that mutual particle avoidance visible in the second order correlation function is more important than the commonly used explanation of broadened single-particle orbitals \cite{Li2006}.

\section{multi-orbital vs. dressed-band description}
\label{SEC:2_alt_description}
We start with the single-particle Hamiltonian describing atoms of mass $m$ in a 3D cubic optical lattice,
\spl{
\label{EQ:sp_Hamiltonian}
\mathcal{H}_{\mbox{\tiny lat}}= \frac{\hat {\bp}^2}{2m} + \int d^{3}r \sum_{d=x,y,z} s  \left( \sin^2({\pi r_i/a} ) - \frac 1 2 \right) \ket{\br}\bra{\br}
}
with the same lattice depth $s$ and spacing $a$ in each dimension. We work in units of the recoil energy $E_r=\frac{1}{2m} \left( \frac{\pi \hbar}{a} \right)^2$. Performing a band structure calculation and Fourier transforming the single-particle Bloch eigenstates leads to a multi-orbital basis of Wannier orbitals \footnote{Choosing the complex phases of the individual Bloch states appropriately is vital for the real-space localization of the resulting Wannier states.}, for which we introduce the bosonic annihilation (creation) operators $a_{i,\alpha}$ ($a_{i,\alpha}^\dag$) at site $i$ and in the band $\alpha=(\alpha_x,\alpha_y,\alpha_z)$.
A short-ranged interaction for two atoms scattering in the $s$-wave channel only at the relevant energy scale can be well approximated by a $\delta$-type contact interaction and characterized completely by the $s$-wave scattering length $a_s$. The interaction strength parameter for the effective contact interaction is given by $g=4 \pi \hbar^2 a_s / m$ and the interaction Hamiltonian can thus be expressed in the multiband Wannier basis as
\spl{
\label{EQ:H_int_contact}
\mathcal{H}_{\mbox{\tiny int}}&= \frac g 2 \int d^3r \, \psi^{\dag}(\br) \psi^{\dag}(\br) \psi(\br) \psi(\br)\\
&=\sum_{\stackrel{\alpha_1,\alpha_2,\alpha_3,\alpha_4}{i_1,i_2,i_3,i_4}}  U_{\alpha_1,\alpha_2,\alpha_3,\alpha_4}^{(i_1,i_2,i_3,i_4)} \,a_{i_1,\alpha_1}^{\dag} a_{i_2,\alpha_2}^{\dag} a_{i_3,\alpha_3}^{\phantom{\dag}} a_{i_4,\alpha_4}^{\phantom{\dag}}
}
where the matrix elements are defined in terms of the single-particle Wannier functions
\begin{equation}
          U_{\alpha_1,\alpha_2,\alpha_3,\alpha_4}^{(i_1,i_2,i_3,i_4)}=\frac g 2 \int d^3r \,w_{i_1,\alpha_1}^*(\br)\,w_{i_2,\alpha_2}^*(\br)\,w_{i_3,\alpha_3}(\br)\,w_{i_4,\alpha_4}(\br)
\end{equation}
Together with the contact interaction term, the full many-body interacting lattice Hamiltonian can be be written in terms of five contributions:
\spl{
\label{EQ:H_all_terms}
\mathcal{H}_{\mbox{\tiny tot}}&=\mathcal{H}_{\mbox{\tiny lat}}+\mathcal{H}_{\mbox{\tiny int}}-\mu \sum_{i,\alpha} a_{i,\alpha}^{\dag} a_{i,\alpha}^{\phantom{\dag}} \\
&= \mathcal{H}_{\epsilon} + \mathcal{H}_{ U,\mbox{\footnotesize loc}} + \mathcal{H}_{t} + \mathcal{H}_{ U,\mbox{\footnotesize nn}}+\mathcal{H}_{\mbox{\footnotesize lr}}
}

Here, $\mathcal{H}_{\epsilon}=\sum_\alpha (\epsilon^{(\alpha)}-\mu) \sum_i a_{i,\alpha}^\dag a_{i,\alpha}^{\phantom{\dag}}$ is the on-site contribution of the single-particle lattice Hamiltonian given by Eq.~(\ref{EQ:sp_Hamiltonian}), with $\epsilon^{(\alpha)}$ being the mean energy of the band $\alpha$ and $\mu$ being the chemical potential when switching to the grand canonical ensemble.
The term $\mathcal{H}_{t} =\sum_\alpha  t^{(\alpha)}\sum_{\langle i,j \rangle} (a_{i,\alpha}^\dag a_{j,\alpha}^{\phantom{\dag}} +\mbox{h.c.})$ is the tunneling between all pairs of nearest-neighboring sites ${\langle i,j \rangle}$ within the different bands $\alpha$. $t^{(\alpha)}=\frac{1}{L}\sum_{\bk}e^{ia\bk \cdot \mathbf e_i}  E^{(\alpha,\bk)}$ is the nearest-neighbor tunneling energy along direction $\mathbf e_i$, i.e. the first component of the energy band $E^{(\alpha,\bk)}$'s Fourier transform, with the sum of quasi-momenta $\bk$ extending over the first Brillouin zone of a lattice containing $L$ sites.

Note that the terms $\mathcal{H}_{\epsilon}, \mathcal{H}_{t}$ and a part of $\mathcal{H}_{\mbox{\footnotesize lr}}$ do not couple different bands, whereas the on-site interaction term $\mathcal{H}_{ U,\mbox{\footnotesize loc}}$ conserves the local many-body parity
\begin{equation}
          Q_i^{(x)}=\prod_{\alpha_x=1,3,5,\ldots} \: \prod_{\alpha_y,\alpha_z=0}^\infty (-1)^{ \hat n_{i,\alpha} }
\end{equation}
(for the $x$-dimension, others are analogous) along each dimension, as shown in the Appendix. The local interacting Hamiltonian at every site can thus be diagonalized in the subspace corresponding to all multiorbital local states with the same parity as the ground state. The resulting eigenstates subsequently constitute an alternative set of basis states for the $n$-particle local Hilbert space.

\begin{figure}[t]
\includegraphics[width=\linewidth]{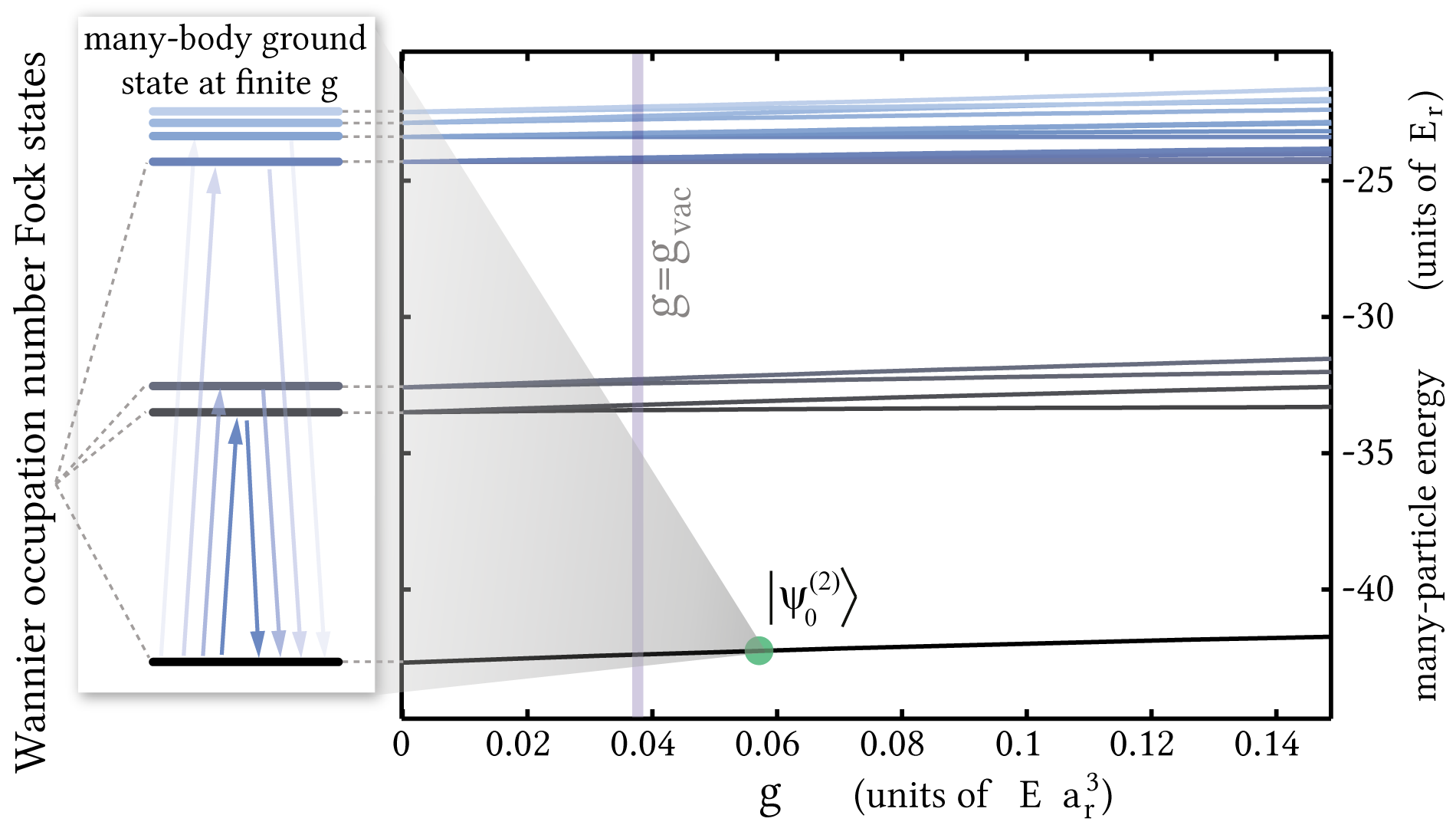}
            \caption
            {\label{fig:spectrum}
(Color online) Two-particle energy spectrum of the local Hamiltonian $\mathcal{H}_{\mbox{\footnotesize loc}}^{(i)}$ in a 3D cubic $768$nm lattice of depth $s=10$ as a function of the interaction strength $g$. For any finite $g$ the local ground state $\ket{\psi_o^{n}}$ is no longer the pure Fock state, with all particles occupying the lowest local Wannier orbital, but an admixture of higher Wannier orbitals is coupled by the local interaction terms. This lowers the total energy (including all orders in perturbation theory) and can be thought of as a dressed state in an effective lowest-band, as it evolves continuously from the $g=0$ limit and remains gapped from all higher dressed bands for typical interaction strengths. $g_{vac}$ is the two-body inter-atomic interaction strength of $^{87}$Rb without the presence of an external magnetic field addressing the Feshbach resonance. In contrast to working in the truncated single orbital Wannier basis, this state gives a much better low-energy description containing local correlations, as was confirmed experimentally \cite{Will10, Will11}. Spatial localization is furthermore guaranteed, since only higher Wannier orbitals at the same site are occupied.
}\end{figure}

We now briefly recapitulate the approximations made in the derivation of the standard Bose-Hubbard model: First, one relies on a strong spatial localization of the single-particle Wannier functions. For a sufficiently strong lattice depth $s$, this justifies taking only nearest-neighbor tunneling as well as only on-site interactions into account and neglecting all others. Second, one assumes that all interband couplings (for any relevant operator) are negligible, thus justifying a truncation to the lowest single-particle band before constructing the many-particle Fock space. The presence of additional terms of the first kind is intrinsic to the problem and cannot be remedied. In a lattice of finite depth and with discrete translational symmetry, the basis states in which all of these couplings would disappear necessarily has spatially completely delocalized basis states (i.e. they are the Bloch Fock states in the absence of interactions), which contradicts the initial goal of finding a spatially localized basis. On the other hand, the problem of interband couplings can be remedied by switching to the basis of local eigenstates. Here, the local interband couplings are contained to infinite order in each local eigenstate and the subsequent coupling between the different lattice sites gives rise to an alternative band structure, which we refer to as dressed bands. This evolution of the local many-particle energy spectrum, where the noninteracting bands continuously evolve into dressed bands is shown in Fig.~\ref{fig:spectrum}.

For a noninteracting system, the Fock state with $\ket{n}_i$ particles occupying the lowest Wannier orbital $\alpha=1$ at site $i$ is the local lowest energy state and therefore is well suited as a basis vector for a low-energy description of the system. In the presence of interactions, $\ket{n}_i$ is no longer the local lowest energy state, although the Fock states still provide a complete basis when taking all bands into account \cite{Larson2009}. However, for the simulation of interacting many-particle systems, one is often interested in a single-band description, and it is of great importance to find the best possible effective single-band basis. From these many-particle basis states, one requires that they
\begin{enumerate}
\item have the highest possible spatial localization, i.e. minimize a localization measure such as the spatial variance of the density profile.
\item contain the local interaction-induced correlations, also lowering the many-particle energy expectation values of these states (evaluated with the full interacting Hamiltonian).
\item possess a well defined local particle number, such that the occupation number representation can be associated with these states.
\item are mutually orthogonal and span the complete low-energy subspace, i.e. formally constitute a basis.
\item recover the standard Bose-Hubbard model in the noninteracting limit.
\end{enumerate}

Having defined these requirements which we impose on an optimized effective single-band basis, the next task is to find a set of such states that fulfill the above requirements.
We propose to use the many-particle eigenstates of the local interacting Hamiltonian $\mathcal{H}_{\mbox{\footnotesize loc}}=\mathcal{H}_{\epsilon} + \mathcal{H}_{ U,\mbox{\footnotesize loc}}$ (which is a direct sum of local Hamiltonians $\mathcal{H}_{\mbox{\footnotesize loc}}^{(i)}$) projected onto the Fock space spanned by the set of all noninteracting Wannier orbitals at a single site $i$. The above criteria are then fulfilled for the following reasons:
\begin{enumerate}
\item Maximal spatial localization \footnote{Clearly, this property competes with the energy minimization; we do not, however, require spatial localization beyond the lattice spacing $a$ for the validity of a discretized lattice model. Therefore, the energy reduction criterion dominates, once spatial localization on the order of the lattice spacing is guaranteed.} carries over from the maximum localization of the single-particle Wannier orbitals at a given site.
\item The multi-orbital, many-particle local ground state by definition minimizes the local energy and contains correlations in the interacting case, where the eigenstates are entangled with respect to the single-particle basis.
\item Since the local truncated interacting Hamiltonian conserves the local particle number
\begin{equation}
[\mathcal \mathcal{H}_{\mbox{\footnotesize loc}}^{(i)}, \sum_\alpha a_{i,\alpha}^\dag a_{i,\alpha}^{\phantom{\dag}}]=0,
\end{equation}
the eigenstates of $\mathcal \mathcal{H}_{\mbox{\footnotesize loc}}^{(i)}$ can all be chosen to have a fixed local particle number. For all typical experimental interaction strengths, the ground state is non-degenerate (thus necessarily possessing a fixed particle number). This allows a clear translation from the initial truncated single-band Wannier occupation into the new dressed band formalism: the initial local Fock state $\ket{n_i}_i$ is formally replaced by the local correlated ground state with $n$ particles $\ket{\psi_0{(n)}}$ at the cost of renormalizing the Bose-Hubbard parameters.
\item Local ground states with different particle number are orthogonal (or can be chosen as such in the case of degeneracy) since they are simultaneously eigenstates of the Hermitian local many-particle Hamiltonian. States at different sites on the other hand are orthogonal, since by construction they only occupy Wannier orbitals at different sites, which are orthogonal on the single-particle level.
\item In the noninteracting limit, the local $n$-particle ground state continuously converges to the local Wannier Fock state, thus recovering this limit.
\end{enumerate}

All longer-range matrix elements (beyond nearest-neighbor) from the lattice as well as the interaction Hamiltonian are contained in the long-range term $\mathcal{H}_{\mbox{\footnotesize lr}}$. These will not be discussed in further detail, since their translation into the effective single-band basis is identical to that of the nearest-neighbor terms, they are however generally smaller in magnitude. The remaining terms (on-site and terms connecting nearest-neighbors) from the interaction Hamiltonian can be classified into four groups: on-site interaction terms forming the local interaction Hamiltonian $\mathcal{H}_{ U,\mbox{\footnotesize loc}}^{(i)}=\sum_{\alpha_1,\alpha_2,\alpha_3,\alpha_4,i}  U_{\alpha_1,\alpha_2,\alpha_3,\alpha_4}^{(i,i,i,i)} \,a_{i,\alpha_1}^{\dag} a_{i,\alpha_2}^{\dag} a_{i,\alpha_3}^{\phantom{\dag}} a_{i,\alpha_4}^{\phantom{\dag}}$, the density-induced single-particle tunneling Hamiltonian $\mathcal{H}_{ U,\mbox{\footnotesize nn}}^J$ containing terms of the form $a_{i,\alpha_1}^{\dag} a_{i,\alpha_2}^{\dag} a_{i,\alpha_3}^{\phantom{\dag}} a_{j,\alpha_4}^{\phantom{\dag}}+h.c.$ (as well as their counterparts under exchange $i \leftrightarrow j$), pair tunneling terms of the form $a_{i,\alpha_1}^{\dag} a_{i,\alpha_2}^{\dag} a_{j,\alpha_3}^{\phantom{\dag}} a_{j,\alpha_4}^{\phantom{\dag}}+h.c.$ in $\mathcal{H}_{ U,\mbox{\footnotesize nn}}^I$, as well as the nearest-neighbor interaction Hamiltonian $\mathcal{H}_{ U,\mbox{\footnotesize nn}}^{\mbox{\footnotesize int}}$ with terms $a_{i,\alpha_1}^{\dag} a_{i,\alpha_2}^{\phantom{\dag}} a_{j,\alpha_3}^{\dag}  a_{j,\alpha_4}^{\phantom{\dag}}+\mbox{h.c.}$ All terms of the latter three types are contained in the nearest-neighbor interaction Hamiltonian $\PlowE \mathcal{H}_{ U,\mbox{\footnotesize nn}} \PlowE=\mathcal{H}_{ U,\mbox{\footnotesize nn}}^J+\mathcal{H}_{ U,\mbox{\footnotesize nn}}^I+\mathcal{H}_{ U,\mbox{\footnotesize nn}}^{\mbox{\footnotesize int}}$.

\section{Definition of the low-energy subspace}
\label{SEC:low_E_subspace}
\begin{figure}[b!]
\includegraphics[width=\linewidth]{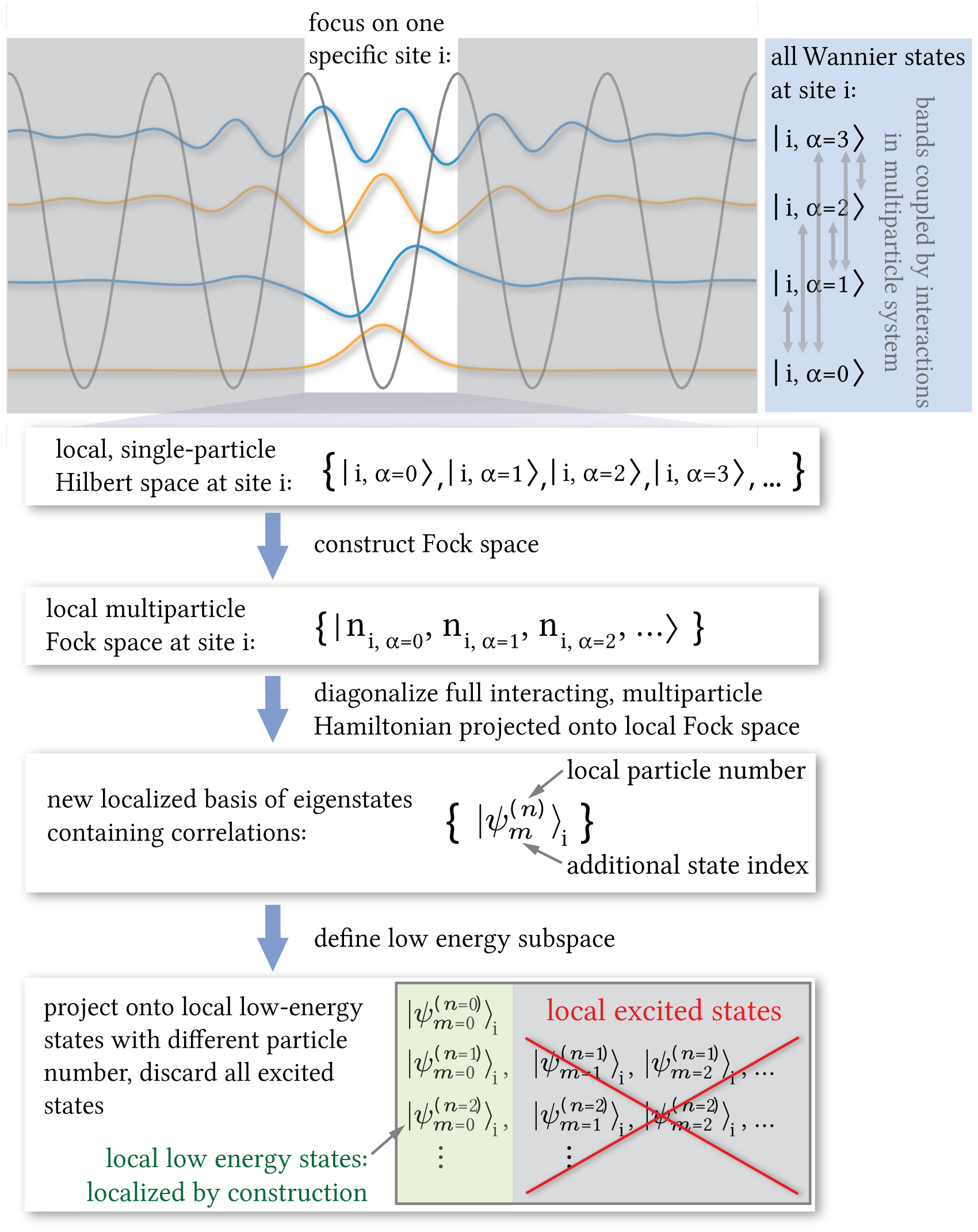}
            \caption
            {\label{fig:procedure}
(Color online) Cartoon depicting our procedure to construct the localized low-energy, dressed-band basis. The Wannier functions (shown for a lattice of depth $s=8E_r$) are obtained from an exact single-particle band structure calculation on a large lattice, before one focuses on the diagonalization of the interacting many-particle Hamiltonian on the Fock subspace of multiorbital maximally localized Wannier states on a single site. This is not to be confused with a truncation of the spatial potential to a single lattice site.
}\end{figure}

In this section, we systematically construct the effective low-energy subspace. The full Hamiltonian projected onto this subspace gives the best possible description of interacting bosons in a lattice at a sufficiently low temperature, where all higher dressed bands can be neglected. Our procedure is summarized in Fig.~\ref{fig:procedure}.

We diagonalize the local part of the Hamiltonian in Eq.~(\ref{EQ:H_all_terms}), which is a direct sum of local Hamiltonians
\spl{
\mathcal{H}_{\mbox{\footnotesize loc}}=\mathcal{H}_{\epsilon} + \mathcal{H}_{ U,\mbox{\footnotesize loc}}=\sum_i \mathcal{H}_{\mbox{\footnotesize loc}}^{(i)} \otimes \prod_{\otimes j\neq i} \mathbbm{1}_j.
}
This can be achieved by diagonalizing the local Hamiltonian at each site separately, although for a homogeneous lattice the diagonalizations are of course identical and it suffices to perform one only. Our formalism is, however, directly extensible to inhomogeneous systems, e.g., in the presence of additional spatial potentials or spatially dependent interactions. Note that in our notation, $\mathcal{H}_{\mbox{\footnotesize loc}}^{(i)}$ operates only on the \emph{local} Hilbert space of a single site (i.e., not on the complete lattice Fock space); the complete many-particle lattice Hilbert space is the direct product of all local Fock spaces over all sites.
Truncating the local single-particle space to the $\bcut$ lowest bands, we diagonalize the local Hamiltonian in the Wannier Fock representation at fixed particle number and parity, leading to
\spl{
\mathcal{H}_{\mbox{\footnotesize loc}}^{(i)}=\sum_{m,n} E_m^{(n)} \ket{\psi_m^{(n)}}_i {_i}\bra{\psi_m^{(n)}}.
}
Due to number conservation of $\mathcal{H}_{\mbox{\footnotesize loc}}^{(i)}$, the eigenstates $\ket{\psi_m^{(n)}}_i$ can always be chosen to be of fixed particle number $n$ and contain an additional excitation index $m$.
On a local level, a projection onto the many-particle low-energy space means only considering the local correlated ground states,
\begin{equation}
\ket{\psi_0^{(n)}}_i = \sum_{n_{i,0}, \ldots, n_{i,\bcut}} c_{n_{i,0}, \ldots, n_{i,\bcut}} \ket{n_{i,0}, \ldots, n_{i,\bcut}}
\end{equation}
with all different particle numbers $n = n_{i,0} + \ldots + n_{i,\bcut}$. This low-energy projection can be extended to the whole system by defining the subspace spanned by the basis states,
\begin{equation}
\label{EQ:many_site_lowE_state}
\ket{\psi_0^{(n_1,\ldots,n_L)}} \equiv \prod_{\otimes i=1}^L \ket{\psi_0^{(n_i)}}_i
\end{equation}
for all possible sets of integer local occupation numbers $(n_1,\ldots,n_L)$. By construction, these states are all mutually orthogonal,
\begin{equation}
   \braket{\psi_0^{(n_1,\ldots,n_L)}}{\psi_0^{(n_1',\ldots,n_L')}}=\delta_{n_1,n_1'} \ldots \delta_{n_L,n_L'},
\end{equation}
which follows from the properties of the direct product in combination with the local states being different eigenstates of the same Hermitian Hamiltonian. It is also useful to define a low-energy projection operator
\begin{align}
\begin{split}
\label{EQ:P_low_E_total}
\PlowE=\sum_{n_1,\ldots,n_L}\ket{\psi_0^{(n_1,\ldots,n_L)}} \bra{\psi_0^{(n_1,\ldots,n_L)}}
\end{split}
\end{align}
which projects any state from the full multi-orbital Fock space to the low-energy subspace of the full lattice.

\section{Transformation into the new dressed-band basis}
\label{SEC:basis_transformation}
Having defined the effective single-band space of interest, we now focus on expressing arbitrary operators in this subspace. Here it proves very useful to define a set of new ladder operators,
\spl{
\label{EQ:new_b_operators}
b_i = \left(  \sum_{n=1}^\infty \sqrt n \ket{\psi_0^{(n-1)}}_i {_i}\bra{\psi_0^{(n)}} \right) \otimes \prod_{\otimes j\neq i}\mathbbm{1}_j
}
where $i$ again refers to a physical site. It can be seen from the structure of these operators that any operator containing only transition elements between low-energy states of the type in Eq.~(\ref{EQ:many_site_lowE_state}) can be expressed in terms of these ladder operators and their Hermitian conjugates. Furthermore, it can be directly verified that these operators fulfill bosonic commutation relations,
\begin{equation}
[ b_i, b_{j}^\dag  ] = \delta_{i,j}.
\end{equation}
Consequently, these ladder operators take over the role of the Wannier orbital creation and annihilation operators within a more appropriate single-band description of an interacting bosonic lattice system. The next step is to express the original Hamiltonian and any other $N$-particle operator in terms of the operators $b_i$ and $b_i^\dag$, after projection onto the lowest dressed band. This does not mean that all Wannier creation and annihilation operators $a_{i,\alpha}$ and $a_{i,\alpha}^\dag$ are directly substituted by $b_i$ and $b_i^\dag$, but a systematic transformation is required, which we will now derive.

\begin{figure}[t!]
\includegraphics[width=\linewidth]{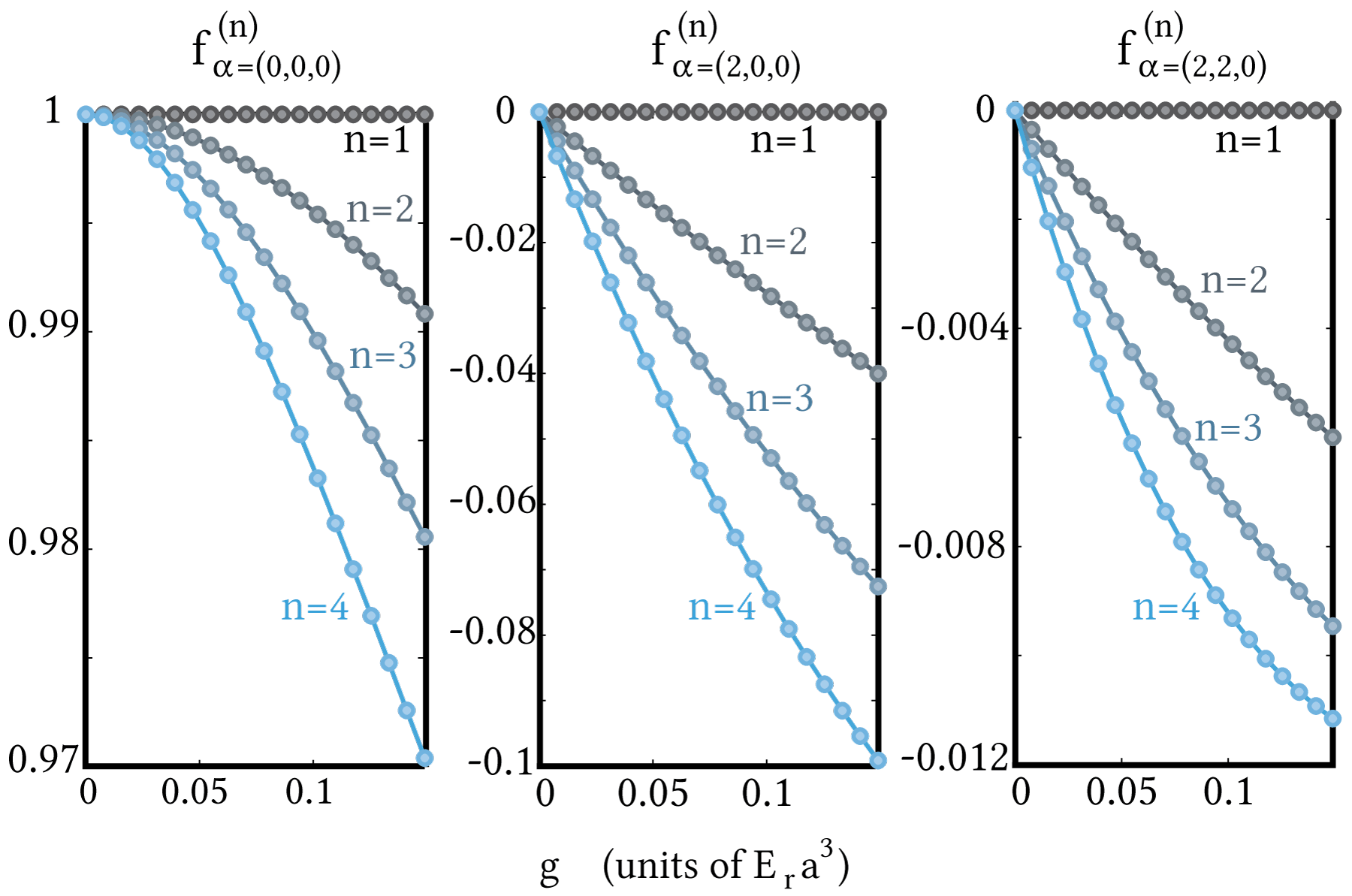}
            \caption
            {\label{fig:f_coeffs}
(Color online) The coefficients $f_\alpha^{(n)}$ as a function of interaction strength $g$ for a $738$nm 3D cubic lattice of depth $s=10E_r$. Note the symmetry relation $f_{\alpha=(2,0,0)}^{(n)} = f_{\alpha=(0,2,0)}^{(n)}=f_{\alpha=(0,0,2)}^{(n)}$ and $f_{\alpha=(2,2,0)}^{(n)} = f_{\alpha=(2,0,2)}^{(n)}=f_{\alpha=(0,2,2)}^{(n)}$ if the lattice is isotropic. In the noninteracting limit where all particles occupy the lowest local Wannier orbital, the coefficients $f_{\alpha=(0,0,0)}^{(n)}$ approach $1$, whereas all coefficients corresponding to other orbitals vanish.
}
\end{figure}

An arbitrary operator (acting on the full lattice) $\mathcal{D}=\sum_l \mathcal{D}^{(l)}$, expressed in terms of multiorbital lattice Wannier operators, can be decomposed into normally ordered terms, where each term $\mathcal{D}^{(l)}$ can contain operators corresponding to many different lattice sites. Projecting this operator onto the lowest dressed band, i.e. multiplying with  operator $\PlowE $ from both the left and right, decouples this operator in the sense that the contribution to each lattice site can be considered individually. Omitting the site index, one such local term is thus generally of the normally ordered form
\begin{equation}
\label{EQ:local_operator}
A^{(i)} =  a_{i,\alpha_1}^\dag \ldots a_{i,\alpha_p}^\dag a_{i,\beta_1} \ldots a_{i,\beta_q},
\end{equation}
containing $p$ creation and $q$ annihilation operators and acting as the unit operator on all other sites. 
We introduce the projector on the low-energy subspace at site $i$, which can be explicitly written as
\spl{
\label{EQ:Plow_E_i}
\PlowE^{(i)}=\sum_{n=0}^\infty \ket{\psi_0^{(n)}}_i {_i}\bra{\psi_0^{(n)}} \otimes \prod_{ \otimes j\neq i} \mathbbm{1}_j.
}
This is related to the projector on the entire low-energy subspace in Eq.~(\ref{EQ:P_low_E_total}) by the operator product over all sites
\spl{
\PlowE=\prod_i \PlowE^{(i)}.
}
In the following, we first concentrate on the transformation of the local operator $A^{(i)}$ at site $i$ only and omit writing the product with the local unity operators at all other sites, which is implied. Equation~(\ref{EQ:Plow_E_i}) can also be seen as the completeness relation within the local low-energy subspace, which we insert twice into Eq.~(\ref{EQ:local_operator}),
\spl{
\label{EQ:Ai_projected_onto_low_E}
 \PlowE^{(i)} A^{(i)}\PlowE^{(i)} &=\sum_{m,n=0}^\infty \ket{\psi_0^{(m)}}_i {_i}\bra{\psi_0^{(m)}} A^{(i)} \ket{\psi_0^{(n)}}_i {_i}\bra{\psi_0^{(n)}}.
}
Since $A^{(i)}$ in Eq.~(\ref{EQ:local_operator}) contains $p$ creation and $q$ annihilation operators, the central matrix element in Eq.~(\ref{EQ:Ai_projected_onto_low_E}) identically vanishes unless $m-p=n-q$ and we have ${_i}\bra{\psi_0^{(m)}} A^{(i)} \ket{\psi_0^{(n)}}_i \propto \delta_{m-p,n-q}$. Together with a prefactor, which will be useful for symmetry properties and a later transformation relation, we define the matrix elements
\spl{
\label{EQ:def_gerneralized_f}
f_{\boldsymbol \alpha_p, \, \boldsymbol \beta_q}^{(n-q+1)} =\frac{(n-q)!}{\sqrt{n!(n+p-q)!}}  \bra{\psi_0^{(n+p-q)}} A^{(i)}  \ket{\psi_0^{(n)}}.
}
Here we defined the vector notation for the set of band indices $\boldsymbol \alpha_p = (\alpha_1, ..., \alpha_p)$ and $\boldsymbol \beta_q = (\beta_1, ..., \beta_q)$. For notational convenience, we have dropped the site index $i$ of which the $f$'s are independent for a homogeneous lattice.  The upper indices are labeled in a fashion, such that $f_{\boldsymbol \alpha_p,\boldsymbol \beta_q}^{(r)}$ is defined and can be non-zero for any integer $r\geq 1$. These coefficients can be directly calculated once the local eigenstates $\ket{\psi_0^{(n)}}$ are obtained from the exact diagonalization of the local Hamiltonian and the dependence of some typical coefficients $f$ on the interaction strength $g$ is shown in Fig.~(\ref{fig:f_coeffs}). It is sufficient to restrict the indices to $q\geq p$, since all other cases are related by conjugation. For the special case that $A^{(i)}$ consists only of annihilation operators (i.e. $p=0$), we define $f_{\boldsymbol \beta_q}^{(r)}=f_{(),\boldsymbol \beta_q}^{(r)}$ for notational convenience. In Fig.~\ref{fig:f_coeffs} the coefficients $f_\alpha^{(n)}$ for a single annihilation operator $p=0$, $q=1$ are shown. These coefficients have the symmetry property
\begin{equation}        f_{\boldsymbol \beta_q,\boldsymbol \alpha_p}^{(r)}=f_{\boldsymbol \alpha_p,\boldsymbol \beta_q}^{{(r)}^*}
\end{equation}
and are furthermore invariant under permutations of indices within each bracket
\begin{equation}        f_{(\sigma(\alpha_1),\ldots,\sigma(\alpha_p)),(\tilde\sigma(\beta_1),\ldots,\tilde\sigma(\beta_q))}^{{(r)}}=f_{(\alpha_1,\ldots,\alpha_p),(\beta_1,\ldots,\beta_q)}^{{(r)}},
\end{equation}
where $\sigma$ and $\tilde \sigma$ are arbitrary permutations from the symmetric group.

Using the definition in Eq.~(\ref{EQ:def_gerneralized_f}), Eq.~(\ref{EQ:Ai_projected_onto_low_E}) becomes
\spl{
\label{EQ:left_side_trafo_fs}
&\PlowE^{(i)} A^{(i)}\PlowE^{(i)}\\
&=\sum_{n=0}^\infty \frac{\sqrt{n!(n+p-q)!}}{(n-q)!} f_{\boldsymbol \alpha_p, \, \boldsymbol \beta_q}^{(n-q+1)} \ket{\psi_0^{(n+p-q)}}_i {_i} \bra{\psi_0^{(n)}}.
}
It can be seen that the operators in Eq.~(\ref{EQ:left_side_trafo_fs}) take a state with $n$ particles in the low-energy manifold and map it onto a state with $n+p-q$ particles, which is, of course, in accordance with the operator $A^{(i)}$ containing $p$ creation and $q$ annihilation operators. In fact, the set of all operators with this property constitutes a subspace of the operator vector space of all local operators on a given site $i$ operating within the lowest dressed band. Since, on the other hand, the dressed-band operators $b_i$ and $b_i^\dag$ themselves reduce or increase the local particle number by exactly one, operating only within the lowest dressed-band manifold, the set of operators $\{ {b_i^\dag}^{p+m-1} {b_{i}^{\phantom{\dag}}}^{q+m-1} \}$ for given $p,q$ and variable integer $m\geq 1$ spans the same operator subspace. Hence, the operator $\PlowE^{(i)} A^{(i)}\PlowE^{(i)}$ can also be expressed as a superposition of these operators for different $m$, implying the relation
\spl{
\label{EQ:trafo_general_h}
\PlowE^{(i)} A^{(i)}\PlowE^{(i)}
=\sum_{m} h_{\boldsymbol \alpha_p,\boldsymbol \beta_q}^{(m)} \, {b_i^\dag}^{p+m-1} {b_{i}^{\phantom{\dag}}}^{q+m-1},
}
with coefficients $h_{\boldsymbol \alpha_p,\boldsymbol \beta_q}^{(m)}$, which are still to be determined. To determine the explicit relation between the $f$ and $h$ coefficients, we evaluate the matrix elements of Eqs.~(\ref{EQ:left_side_trafo_fs}) and \ref{EQ:trafo_general_h} in the local low-energy basis $\bra{\psi_0^{(n')}}$ and $\ket{\psi_0^{(n)}}$. Here, we keep in mind that, by construction, the operators ${b_i^\dag}$ and ${b_i}$ simply act as ladder operators between the local $n$-particle low-energy states and note that only elements with $n'=n+p-q$ can be non-zero. For each fixed set of band indices $\boldsymbol \alpha_p$ and $\boldsymbol \beta_q$, this leads to the linear transformation relation, which can be written in matrix form as
\spl{
\label{EQ:general_f_g_trafo}
f_{\boldsymbol \alpha_p,\boldsymbol \beta_q}^{(r)} = \sum_{m=1}^\infty \mathcal{B}_{r,m} h_{\boldsymbol \alpha_p,\boldsymbol \beta_q}^{(m)}
}
with the matrix
\begin{equation}
\label{EQ:B_matrix}
\mathcal{B}_{n,m} = \frac{(n-1)!}{(n-m)!} \Theta(n-m)
\end{equation}
and $\Theta(x)$ being the step function with $\Theta(0) = 1$. Explicitly, the first truncated part of the matrix is of the form
\spl{
\mathcal{B}=\left(\begin{tabular}{ c c c c c c c c }
1&0&0&0&0&0&0\\
1&1&0&0&0&0&0\\
1&2&2&0&0&0&0\\
1&3&6&6&0&0&0&\ldots\\
1&4&12&24&24&0&0\\
1&5&20&60&120&120&0\\
1&6&30&120&360&720&720\\
&&&\vdots &&& & \\
\end{tabular}\right)
}
The lower triangular form of the matrix is a consequence of the fact that the operator $(b^\dag)^{p+m} (b)^{q+m}$ annihilates any state with less than $(q+m)$ particles. In the noninteracting case, when the $n$-particle ground state is simply the state with all particles occupying the lowest Wannier orbital $\alpha=0$, we have $f_{\boldsymbol \alpha_p,\boldsymbol \beta_q}^{(r)} = \delta_{\alpha_1,0} \ldots \delta_{\alpha_p,0} \delta_{\beta_1,0} \ldots \delta_{\beta_q,0}$ and $h_{\boldsymbol \alpha_p,\boldsymbol \beta_q}^{(r)} = \delta_{r,1} \delta_{\alpha_1,0} \ldots \delta_{\alpha_p,0} \delta_{\beta_1,0} \ldots \delta_{\beta_q,0}$, in which case the density-induced transitions between the local ground states of different particle number vanish and the effective low-energy creation and annihilation operators are identical to the Wannier creation and annihilation operators.

To explicitly calculate the density-induced transition parameters $h_{\boldsymbol \alpha_p,\boldsymbol \beta_q}^{(m)}$ for a given set of $f$ coefficients, we require the inverse matrix. This is found to be
\begin{equation}
\label{Eq:inverse_B}
(\mathcal{B}^{-1})_{m,n}=\frac{(-1)^{m+n}}{(n-1)! \, (m-n)! } \Theta(m-n)
\end{equation}
with the first elements explicitly being
\spl{ \label{EQ:inverse_Bmatrix}
\mathcal{B}^{-1}=
\left(\begin{array}{ c c c c c c c c}
1&0&0&0&0&0&0\\
-1&1&0&0&0&0&0\\
\frac{1}{2}&-1&\frac{1}{2}&0&0&0&0\\
-\frac{1}{6}&\frac{1}{2}&-\frac{1}{2}&\frac{1}{6}&0&0&0&\ldots\\
\frac{1}{24}&-\frac{1}{6}&\frac{1}{4}&-\frac{1}{6}&\frac{1}{24}&0&0\\
-\frac{1}{120}&\frac{1}{24}&-\frac{1}{12}&\frac{1}{12}&-\frac{1}{24}&\frac{1}{120}&0\\
\frac{1}{720}&-\frac{1}{120}&\frac{1}{48}&-\frac{1}{36}&\frac{1}{48}&-\frac{1}{120}&\frac{1}{720}\\
&&& \vdots
\end{array}\right).
}
It should be pointed out that due to this structure, the truncated inverse matrix is identical to the inverse truncated matrix. This finally allows us to express $\PlowE A^{(i)} \PlowE$ in terms of effective low-energy creation and annihilation operators, which explicitly reads
\spl{
\label{EQ:A_in_terms_of_b_2}
& \PlowE a_{\alpha_1}^\dag ... a_{\alpha_p}^\dag a_{\beta_1} ... a_{\beta_q} \PlowE = f_{\boldsymbol \alpha_p \boldsymbol \beta_q}^{(1)} \, (b^\dag)^p (b)^q \\
& + \biggl( -f_{\boldsymbol \alpha_p \boldsymbol \beta_q}^{(1)} + f_{\boldsymbol \alpha_p \boldsymbol \beta_q}^{(2)} \biggr) (b^\dag)^{p+1} (b)^{q+1} \\
& + \left( \frac 1 2 f_{\boldsymbol \alpha_p \boldsymbol \beta_q}^{(1)} - f_{\boldsymbol \alpha_p \boldsymbol \beta_q}^{(2)} + \frac 1 2 f_{\boldsymbol \alpha_p \boldsymbol \beta_q}^{(3)} \right) (b^\dag)^{p+2} (b)^{q+2} + \ldots.
}
Note that in the noninteracting limit, we have $f_{\boldsymbol \alpha_p \boldsymbol \beta_q}^{(r)} = \delta_{\boldsymbol \alpha_p, \boldsymbol 0} \, \delta_{\boldsymbol \beta_q, \boldsymbol0}$, all coefficients in brackets in Eq.~\ref{EQ:A_in_terms_of_b_2} of higher-order terms cancel, and all operators $a_{i,\alpha}$ ($a_{i,\alpha}^\dag$) for the lowest band can directly be replaced with $b_i$ ($b_i^\dag$).

For the specific case of a single annihilation operator $\PlowE a_\alpha \PlowE$, the transformation is given by
\spl{
& \PlowE a_\alpha \PlowE = f_{\alpha}^{(1)} \, b + \biggl( -f_{\alpha}^{(1)} + f_{\alpha}^{(2)} \biggr) b^\dag b b \\
& + \left( \frac 1 2 f_{\alpha}^{(1)} - f_{\alpha}^{(2)} + \frac 1 2 f_{\alpha}^{(3)} \right) b^\dag b^\dag b b b + \ldots
}
for which we show the first coefficients $f_{\alpha}^{(r)}$ in Fig.~(\ref{fig:f_coeffs}) as a function of the interaction strength. We point out that the local particle number operator transforms as
\begin{eqnarray}
\sum_\alpha a_\alpha^\dag a_\alpha = \sum_{r=1}^\infty \left[ \sum_\alpha h_{(\alpha)(\alpha)}^{(r)} \right] (b^\dag)^r (b)^r = b^\dag b ,
\end{eqnarray}
since $\sum_\alpha f_{(\alpha)(\alpha)}^{(r)} = 1$ for all $r$, and the property of the transformation matrix in Eq.~(\ref{Eq:inverse_B}), $\sum_n (\mathcal{B}^{-1})_{m,n}=\delta_{m,1}$. Hence, the local particle operator in our single dressed band counts all particle in all local orbitals.

The transformation relations are the central result of this section: given any physical operator in the single-particle Wannier basis (as is usually the case) which acts on the system or is measured, one cannot simply substitute the Wannier creation and annihilation operators $a_{i,\alpha}$ with the $b_i$ operators from the effective single-band model with density-dependent parameters. Rather, the transformation relations from Eqns.~(\ref{EQ:trafo_general_h}) and (\ref{EQ:A_in_terms_of_b_2}) have to be used to systematically express this operator in the effective low-energy subspace.

In the following sections, we will apply this transformation to the various terms of the original Hamiltonian. Conceptually, the same transformation is performed on each lattice site. In an inhomogeneous system, the transformations generally differ on different lattice sites, i.e., the coefficients $f$ and $h$ become site dependent. Under the transformation of the full many-body Hamiltonian in Eq.~(\ref{EQ:H_all_terms}), the purely local terms exactly recover the effective multibody interactions found in \cite{Johnson09} and are diagonal in the dressed-band basis with density-dependent interaction parameters. Nonlocal nearest-neighbor terms originate from $\mathcal{H}_{t} + \mathcal{H}_{ U,\mbox{\footnotesize nn}}$ and are non-diagonal in the new basis, leading to the low-energy representation
\spl{\PlowE \mathcal{H}_{\mbox{\tiny tot}} \PlowE&=\PlowE\mathcal{H}_{\mbox{\footnotesize loc}} \PlowE + \mathcal{H}_J + \mathcal{H}_{ U,\mbox{\footnotesize nn}}^I +\mathcal{H}_{ U,\mbox{\footnotesize nn}}^{int}\\&+\PlowE \mathcal{H}_{\mbox{\footnotesize lr}} \PlowE
}
of the full initial Hamiltonian. We neglect all long-range (beyond nearest-neighbor) terms contained in $\mathcal{H}_{\mbox{\footnotesize lr}}$ in this work; their transformation is, however, identical to that of the nearest-neighbor terms. Thereafter, we will discuss an equivalent formulation of the extension of the Bose-Hubbard model using density-dependent parameters for the various terms at the cost of additionally summing over the set of all local low-energy states.

\section{Application to the Bose-Hubbard model: multibody-induced tunneling and interactions}
\label{SEC:multibody_picture}
In this section, we discuss the four relevant terms in the dressed single-band Bose-Hubbard model. These contain all relevant local and nearest-neighbor processes. The amplitudes for processes on neighboring sites in the multibody-induced picture are shown in Fig.~(\ref{FIG:MP_induced_parameters}).

\subsection{Single-particle tunneling term}
\label{SEC:single_particle_tunneling}

\begin{figure*}[t]
\includegraphics[width=\linewidth]{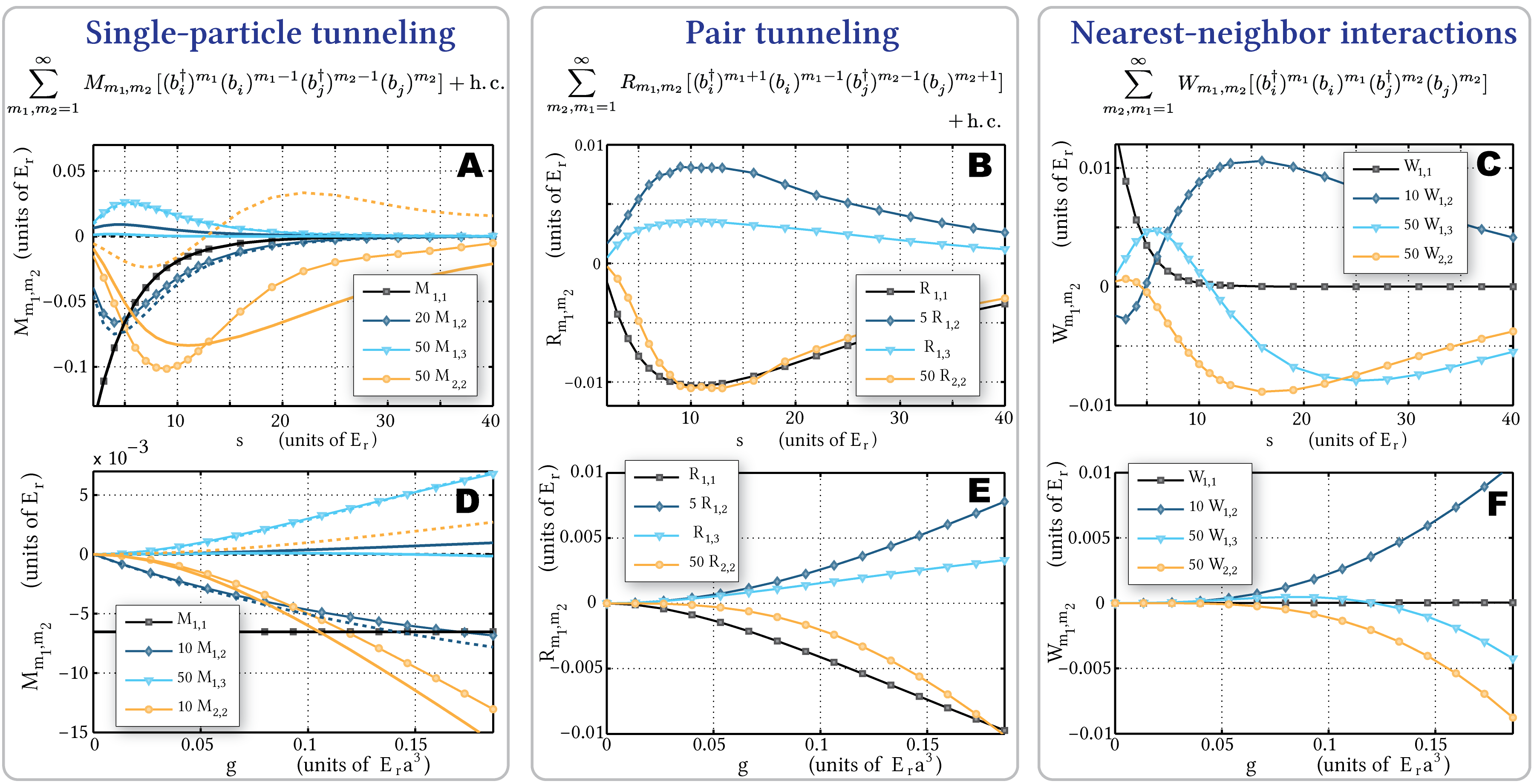}
        \caption
        {\label{FIG:MP_induced_parameters}
        (Color online) Amplitudes for all multibody-induced transitions on nearest-neighboring sites. The lowest order, most relevant processes are shown as functions of the lattice depth $s$ at a fixed interaction strength $g=5 g_{\mbox{\tiny vac}} = 0.186 E_r a^3$ for $^{87}$Rb in the upper row, as well as their dependence on the interaction strength $g$ at a fixed lattice depth $s=10 E_r$ in the lower row. Note the different scaling of individual graphs for visual clarity, which is given in the legends.  The subplots (A,D) in the left column show the effective single-particle tunneling, the dotted and solid lines show the contributions from the interaction term $\mathcal{H}_{ U,\mbox{\footnotesize nn}}$ and the single-particle tunneling term $\mathcal{H}_t$ respectively. With increasing order ($m_1,m_2$) the contribution from the interaction becomes more important and eventually dominates. Whereas the dependence of all three amplitudes on the interaction strength $g$ is monotonic, the dependence on the lattice depth is more complicated, being non-monotonic and even leading to sign changes. We point out that as a function of the lattice depth $s$, the magnitude of the higher-order terms is most significant in the region of $s\approx 10 E_r$ or slightly below, which is also the relevant region for the superfluid-Mott insulator transition (depending on the interaction strength $g$, i.e. the Feshbach resonance). In the noninteracting limit $g\to 0$ all terms, except the lowest order single-particle tunneling $M_{1,1}$ vanish, recovering the usual lowest band Bose-Hubbard model. However, at any finite interaction strength with other terms becoming non-zero, the Bose-Hubbard model truncated to the lowest single-particle Bloch band does not give the correct low-energy description.  The numerical calculations were performed using $6$ bands per dimension, amounting to $216$ single-particle orbitals.
}
\end{figure*}
We now have to gather all operator contributions in the total Hamiltonian $\mathcal{H}_{\mbox{\tiny tot}}$ that give rise to single-particle tunneling transitions between nearest-neighboring sites $i$ and $j$. Clearly, the tunneling term from the original single-particle lattice Hamiltonian $\mathcal{H}_t$ is such a term, but also the  interaction Hamiltonian $\mathcal{H}_{ U,\mbox{\footnotesize nn}}$ contains single-particle transition terms of this type, which we denote by $\mathcal{H}_{U,\mbox{\footnotesize nn}}^J$. For each fixed set of nearest-neighbor sites $i$ and $j$ there are four relevant terms for this process, and thus the total single-particle hopping Hamiltonian originating from the interaction term is
\spl{
\mathcal{H}_{ U,\mbox{\footnotesize nn}}^J=\sum_{\langle i,j\rangle} \sum_{\alpha_1,\alpha_2,\alpha_3,\alpha_4} [ U_{\alpha_1,\alpha_2,\alpha_3,\alpha_4}^{(i,i,i,j)}
\,a_{i,\alpha_1}^{\dag} a_{i,\alpha_2}^{\dag} a_{i,\alpha_3}^{\phantom{\dag}} a_{j,\alpha_4}^{\phantom{\dag}}
\\+ U_{\alpha_1,\alpha_2,\alpha_3,\alpha_4}^{(i,j,j,j)}
\,a_{i,\alpha_1}^{\dag} a_{j,\alpha_2}^{\dag} a_{j,\alpha_3}^{\phantom{\dag}} a_{j,\alpha_4}^{\phantom{\dag}}] + \mbox{h.c.}
}
Note that in contrast to $\mathcal{H}_t$, which is diagonal in the band index, $\mathcal{H}_{ U,\mbox{\footnotesize nn}}^J$ couples Wannier states in different bands on neighboring sites.

The single-particle tunneling $\mathcal{H}_J$ can also directly be written in terms of $b$-operators, giving rise to multiparticle-induced single-particle tunneling,
\spl{
\label{EQ:SP_eff_tunneling}
\mathcal{H}_J&=\PlowE\mathcal{H}_t \PlowE+\mathcal{H}_{U,\mbox{\footnotesize nn}}^J\\
&= \sum_{m_1,m_2=1}^\infty M_{m_1,m_2} \sum_{\langle i,j \rangle} [ (b_{i}^\dag )^{m_1} (b_{i}^{\phantom{\dag}})^{m_1-1}(b_{j}^\dag )^{m_2-1} (b_{j}^{\phantom{\dag}})^{m_2} ] +\mbox{h.c.}
}
with the $(m_1,m_2)$-particle-induced tunneling amplitude
\spl{
 M_{m_1,m_2}=\sum_\alpha  t^{(\alpha)} {h_\alpha^{(m_1)}}^* {h_\alpha^{(m_2)}}+   \sum_{\alpha_1,\alpha_2,\alpha_3,\alpha_4} \left[ U_{\alpha_1,\alpha_2,\alpha_3,\alpha_4}^{(i,i,i,j)} \right. \\ \times \left. h_{(\alpha_3)(\alpha_2 \alpha_1)}^{{(m_1-1)}^*}  {h_{\alpha_4}^{(m_2)}} +  U_{\alpha_1,\alpha_2,\alpha_3,\alpha_4}^{(i,j,j,j)}  {h_{(\alpha_2)(\alpha_3 \alpha_4)}^{(m_2-1)}}  h_{\alpha_1}^{{(m_1)}^*} \right],
}
shown in Fig.~\ref{FIG:MP_induced_parameters} A and D. For $m_1=m_2=1$ Eq.~(\ref{EQ:SP_eff_tunneling}) is simply a usual single-particle tunneling term, containing all multi-orbital contributions in $\mathcal{H}_t$, but no contribution from $\mathcal{H}_{U,\mbox{\footnotesize nn}}^J$, since ${h_{(\alpha_1)(\alpha_2 \alpha_3)}^{(m_2)}}$ vanishes for any $m_2<1$. However, there are also additional \emph{multibody-induced} single-particle tunneling terms present: for $m_2>1$ or $m_1>1$ a single-particle can tunnel between neighboring lattice sites with an amplitude $M_{m_1,m_2}$ if $m_2-1$ and $m_1-1$ \emph{additional} particles (additional to the one tunneling) are present on the lattice sites. We therefore refer to these processes as being multibody-induced.

\subsection{Two-particle correlated hopping}
\label{SEC:two_particle_tunneling}
In contrast to the noninteracting lattice Hamiltonian, the interaction term $\mathcal{H}_{ U,\mbox{\footnotesize nn}}$ also contains two-particle correlated tunneling transition elements in the term $\mathcal{H}_{ U,\mbox{\footnotesize nn}}^I$. A single application of such a term to the state $\ket{\psi_0^{(n_i)}}_i\ket{\psi_0^{(n_j)}}_j$ leads to a correlated tunneling of two particles on neighboring sites $i$ and $j$, leading to states of the form $\ket{\psi_0^{(n_i+2)}}_i\ket{\psi_0^{(n_j-2)}}_j$. Clearly such operator terms are beyond the standard Bose-Hubbard model and cannot be contained in a renormalized tunneling parameter. They may however lead to interesting effects and we additionally include them in an extended description of the interacting lattice model.
\spl{
\label{EQ:2P_eff_tunneling}
\mathcal{H}_{ U,\mbox{\footnotesize nn}}^I&= \sum_{m_1,m_2=1}^\infty R_{m_1,m_2} \sum_{\langle i,j \rangle} [ (b_{i}^\dag )^{m_1+1} (b_{i}^{\phantom{\dag}})^{m_1-1}\\
&\times (b_{j}^\dag )^{m_2-1} (b_{j}^{\phantom{\dag}})^{m_2+1} ] +\mbox{h.c.}
}
with the $(m_1,m_2)$-particle-induced two-particle tunneling amplitude
\spl{
R_{m_1,m_2}= \sum_{\alpha_1,\alpha_2,\alpha_3,\alpha_4}  U_{\alpha_1,\alpha_2,\alpha_3,\alpha_4}^{(i,i,j,j)}  {h_{\alpha_1,\alpha_2}^{(m_1)^*}} {h_{\alpha_3,\alpha_4}^{(m_2)}}.
}
The lowest order processes are shown as a function of the lattice depth $s$ and interaction strength $g$ in Fig.~\ref{FIG:MP_induced_parameters} B and E. The magnitude of these amplitudes decrease with increasing order ($m_1, m_2$) of the processes. At large $s$ however, the two-particle tunneling amplitudes decay much slower than the bare single-particle tunneling $J$, such that these processes become relevant on this nearest-neighbor energy scale.

\subsection{nearest-neighbor interactions}
\label{SEC:nn_interactions}

Counting terms corresponding to nearest-neighbor interactions in the total nearest-neighboring part of the Hamiltonian $\sum_{\alpha_1,\alpha_2,\alpha_3,\alpha_4}  U_{\alpha_1,\alpha_2,\alpha_3,\alpha_4}^{(i_1,i_2,i_3,i_4)} \,a_{i_1,\alpha_1}^{\dag} a_{i_2,\alpha_2}^{\dag} a_{i_3,\alpha_3}^{\phantom{\dag}} a_{i_4,\alpha_4}^{\phantom{\dag}}$ with all $i_m$ being one of two nearest-neighbor sites $i$ and $j$, there are four terms corresponding to nearest-neighbor interactions. These are all equivalent and after permuting indices the full nearest-neighbor interaction Hamiltonian can be written in the form
\spl{
\label{EQ:NN_int_induced}
\mathcal{H}_{ U,\mbox{\footnotesize nn}}^{int}&= 4\sum_{\langle i,j \rangle} \sum_{\alpha_1,\alpha_2,\alpha_3,\alpha_4}  U_{\alpha_1,\alpha_2,\alpha_3,\alpha_4}^{(i,i,j,j)}\\ &\times\,\PlowE \, a_{i,\alpha_1}^{\dag} a_{i,\alpha_2}^{\phantom{\dag}}a_{j,\alpha_3}^{\dag} a_{j,\alpha_4}^{\phantom{\dag}}\PlowE\\
&=\sum_{m_1,m_2=1}^\infty \sum_{\langle i,j \rangle} W_{m_1,m_2} \, (b_{i}^\dag )^{m_1} (b_{i}^{\phantom{\dag}})^{m_1}(b_{j}^\dag )^{m_2} (b_{j}^{\phantom{\dag}})^{m_2},
}
where the last line is in the effective multibody nearest-neighbor interaction picture with the parameters
\spl{
W_{m_1,m_2}=4\sum_{\alpha_1,\alpha_2,\alpha_3,\alpha_4}  U_{\alpha_1,\alpha_2,\alpha_3,\alpha_4}^{(i,i,j,j)} {h_{(\alpha_1)(\alpha_2)}^{(m_1)}}\, {h_{(\alpha_3)(\alpha_4)}^{(m_2)}}.
}
These are shown in Fig.~\ref{FIG:MP_induced_parameters} C and F. Note that in contrast to the tunneling Hamiltonians, Eq.~(\ref{EQ:NN_int_induced}) does not contain the addition of the Hermitian conjugate, since this is equivalent to the respective process itself and leads to a factor of $2$.

\subsection{On-site terms}
After diagonalization, we explicitly have the Hamiltonian containing all local terms in diagonal form
\begin{equation}
\label{EQ:H_loc_diag}
\PlowE \mathcal{H}_{\mbox{\footnotesize loc}} \PlowE = \sum_i\sum_n E_0^{(n)}  \ket{\psi_0^{(n)}}_i {_i}\bra{\psi_0^{(n)}} .
\end{equation}
In the density-induced picture containing effective higher-order interaction terms, we seek a representation in terms of the $b$ operators. Since by construction this is diagonal in the lowest dressed-band basis, it can be written as a series of local terms, each containing the same number of creation and annihilation operators
\begin{equation}
\label{EQ:H_U_loc_in_b}
\PlowE \mathcal{H}_{ U,\mbox{\footnotesize loc}}  \PlowE = \sum_i (\epsilon^{(0)} - \mu) b_i^\dag b_i + \sum_i \sum_{m=2}^\infty \frac{V_m}{m!} (b_i^\dag)^m (b_i)^m.
\end{equation}
\begin{figure}[t]
\includegraphics[width=0.9\linewidth]{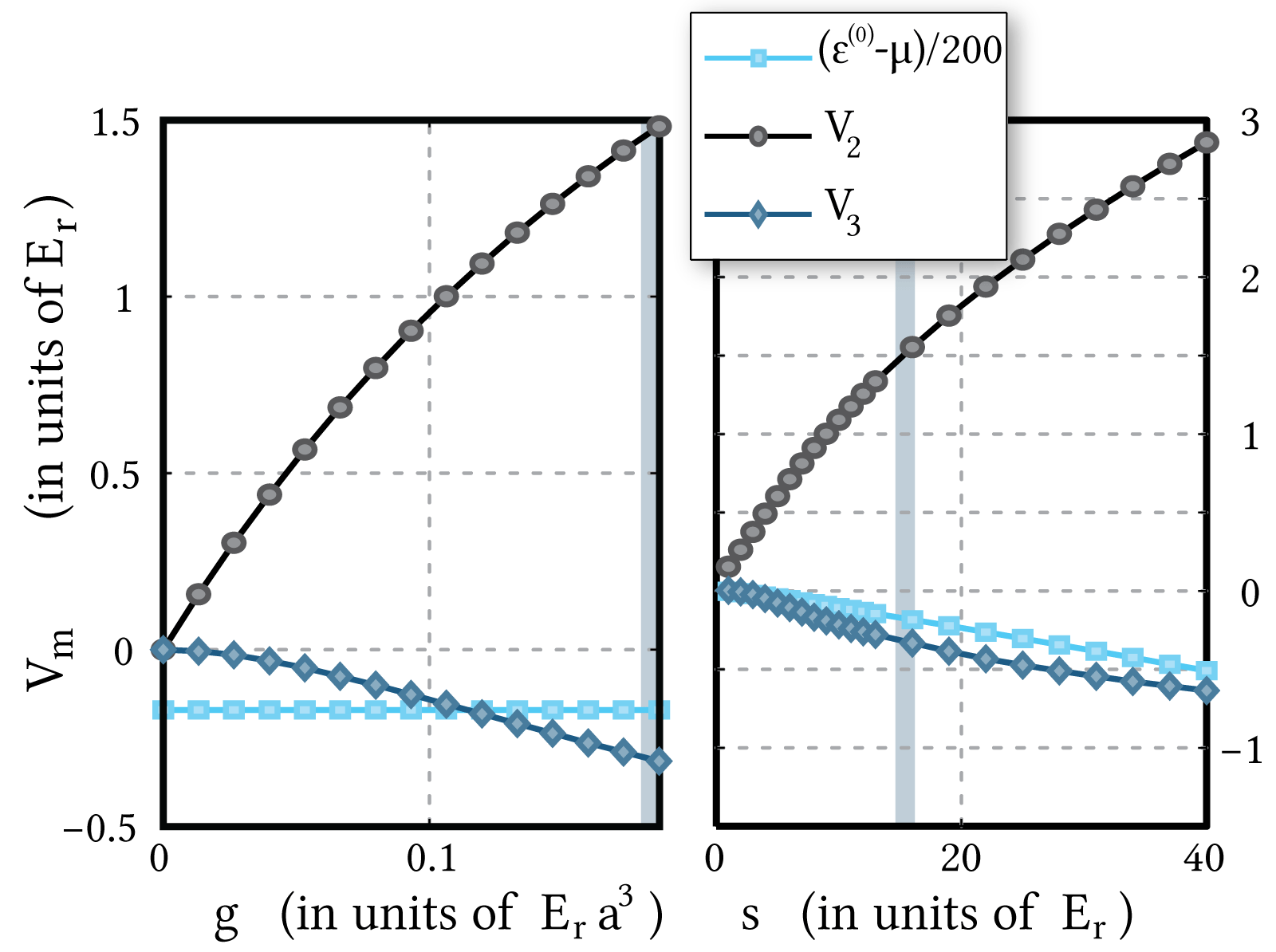}
        \caption
        {\label{FIG:MP_induced_local}
(Color online) The density-induced local parameters as a function of the interaction strength $g$ at constant $s=15 E_r$ (left) and as a function of the lattice depth $s$ at constant $g=5 g_{vac}$ (right). The single-particle term $(\epsilon^{(0)}-\mu)$ coupling to the local density operator is not invariant under a single-particle energy shift. All higher-order terms $V_m$ with $m \geq 2$ are invariant under such a transformation. The two gray shaded areas correspond to the respective region of the other plot.}
\end{figure}

The first term is the single-particle contribution and contains the energy offset of every particle due to the on-site Wannier energy and the chemical potential. It only contains the single-particle lowest band energy, since for the case of a single local particle, interactions do not play a role and the local state $\ket{\psi_0^{(n=1)}}_i$ is simply the lowest band Wannier state. Letting the two equations (\ref{EQ:H_U_loc_in_b}) and (\ref{EQ:H_loc_diag}) act on the local low-energy basis states $\ket{\psi_0^{(n)}}_i$ for all integer $n$ leads to the expression of the higher-order interaction amplitudes in terms of on-site many-particle eigenenergies
\begin{equation}
V_m = m! \sum_{n=1}^\infty (\mathcal{B}^{-1})_{m,n} \frac{E_0^{(n)}}{n}.
\end{equation}
These terms are exactly the effective many-body interactions introduced in \cite{Johnson09} and experimentally observed in \cite{Will10}. They gain significance with both increasing lattice depth $s$ and interaction strength $g$, as shown in Fig.~\ref{FIG:MP_induced_local}.

\section{Density-dependent parameter formulation of the Bose-Hubbard model}
\label{SEC:density_dependent_parameters}
\begin{figure*}[t]
\includegraphics[width=\linewidth]{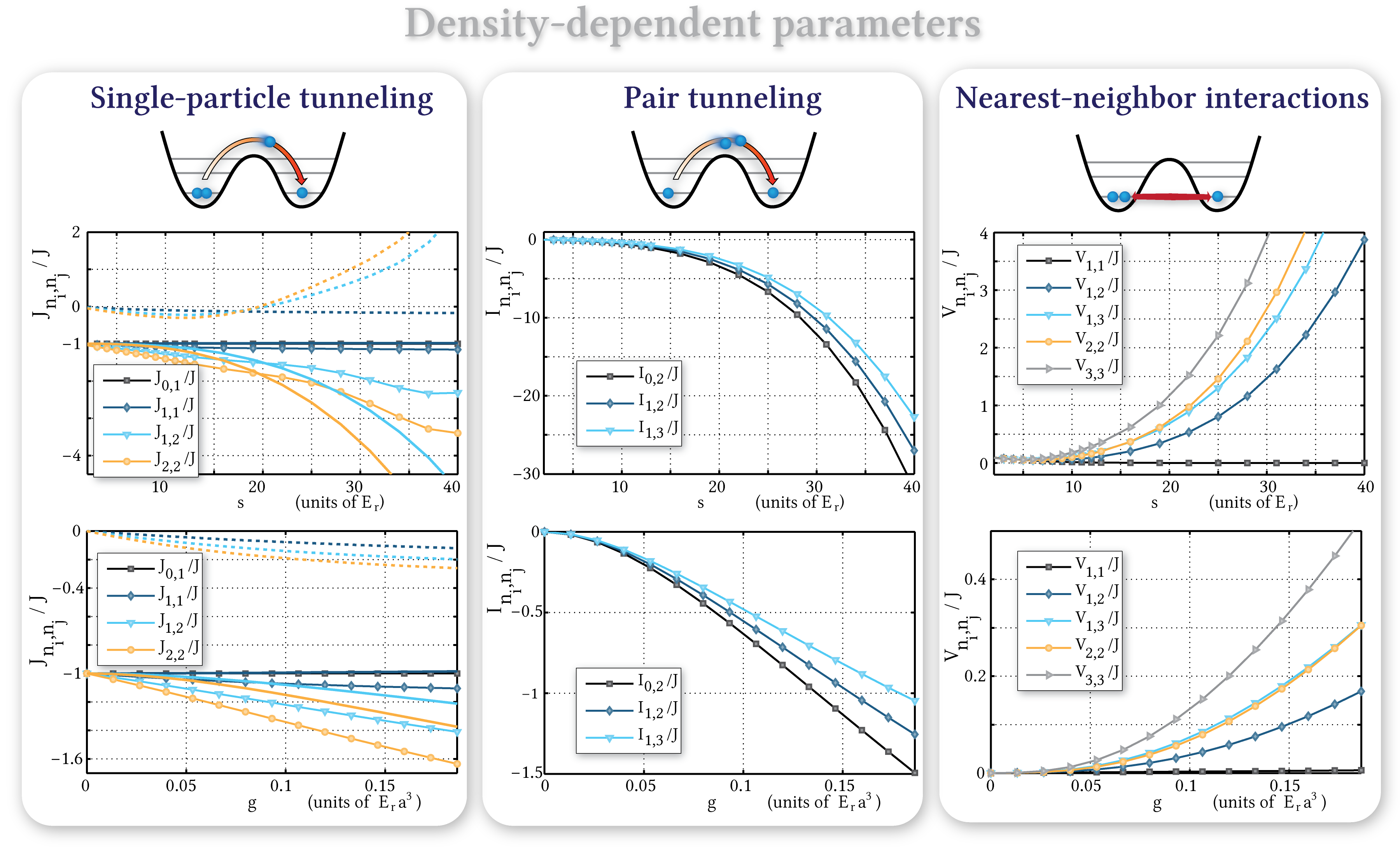}
        \caption
        {\label{FIG:density_dependent_comb}
(Color online) Density-dependent parameters in the lowest dressed band as a function of the lattice depth $s$ at fixed $g=5 g_{vac}$ (upper subfigures A,B,C) and as a function of the interaction strength $g$ at fixed $s=15E_r$ (lower subfigures D,E,F). For the single-particle tunneling terms (subplots A and D) the solid lines indicate the contribution from the multi-orbital single-particle lattice Hamiltonian, whereas the dashed lines are the contributions from the interacting Hamiltonian $\mathcal{H}_{\mbox{\tiny int}}$ respectively. The relevant energy scale for nearest-neighbor processes is set by the bare single-particle tunneling $J=-J_{0,1}$ by which we scale all other quantities in this figure. Note that for strong lattice depths the two-particle tunneling amplitudes and nearest-neighbor interaction energies, which are usually neglected, become relevant and can even dominate on the relevant energy scale $J$. In the noninteracting limit and approximately in the limit of shallow lattice depths, the single-particle tunneling amplitudes become density-independent and approach the single bare particle tunneling amplitude, whereas the pair tunneling amplitudes (subplots B and E) and nearest-neighbor interaction energies (subplots C and F) vanish.
}
\end{figure*}
An alternative and equivalent description to the multibody-induced picture is a formulation in terms of density-dependent parameters. In this picture, one performs a summation over all local occupation numbers and allows the matrix elements, which are coefficients of operators of the form $\ket{\psi_0^{n_i}}_i {_i}\bra{\psi_0^{n_i'}}$ to acquire a density dependence beyond the bosonic statistical factor. This picture is convenient to directly infer certain multiorbital effects, such as the observed energy peak positions in quantum phase revival spectroscopy experiments \cite{Will10, Will11}. However, the density-dependent representation is not always the most convenient approach for treating multiorbital effects with usual many-body methods. For instance, bosonization or even the site decoupling mean-field theory cannot be performed within this framework. The effective representation discussed in Sec.~(\ref{SEC:multibody_picture}), which does not require an external summation over all local occupation numbers is more favorable in this sense and has been successfully applied for the pure on-site terms in describing effective multibody interactions \cite{Johnson09}. We show that, especially in the regime of deep lattices, the density-dependent parameters are strongly modified.  Two-particle tunneling and nearest-neighbor interaction processes, beyond the usual Bose-Hubbard model, become  significant, even becoming an order of magnitude stronger than the bare single-particle hopping $J$ in certain experimentally accessible regimes.

\subsection{Single-particle tunneling term}
\label{SEC:dd_single_particle_tunneling}

We first seek the single-particle tunneling term in the Hamiltonian with a density-dependent tunneling parameter, i.e. of the form
\spl{
\label{EQ:HJ_density_dep}
\mathcal{H}_J=&\sum_{\langle i,j\rangle} \sum_{n_i,n_j} J_{n_i,n_j} \sqrt{n_i+1}\, \ket{\psi_0^{(n_i+1)}}_i {_i}\bra{\psi_0^{(n_i)}}\\
& \otimes  \sqrt{n_j}\, \ket{\psi_0^{(n_j-1)}}_j {_j}\bra{\psi_0^{(n_j)}} + \mbox{h.c.}
}
In a recent independent calculation by L\"uhmann et al. \cite{Luehmann11}, similar results to the ones presented here were obtained for the density-dependent single-particle tunneling parameters within the fully correlated many-body framework.
In this section we omit writing the unit operator on other sites for any local operator: i.e. any operator $A^{(i)}$ acting only on the local Fock space of site $i$ is to be implicitly understood as being extended to the full lattice Fock space as $A^{(i)}\otimes \prod_{\otimes j\neq i} \mathbbm{1}_j$. Each operator term in Eq.~(\ref{EQ:HJ_density_dep}) can also be written with the use of local projectors $\mathcal{P}_n^{(i)}=\ket{\psi_0^{(n)}}_i {_i}\bra{\psi_0^{(n)}}$ on the $n$-particle ground state at site $i$ as
\spl{
 \sqrt{n_i+1}\, \ket{\psi_0^{(n_i+1)}}_i {_i}\bra{\psi_0^{(n_i)}} & \otimes  \sqrt{n_j}\, \ket{\psi_0^{(n_j-1)}}_j {_j}\bra{\psi_0^{(n_j)}}\\=b_i^\dag \, b_j \,\mathcal{P}_{n_i}^{(i)} \mathcal{P}_{n_j}^{(j)}.
}
The final summation over all occupation numbers $n_i,n_j$ is however always necessary in this density-dependent parameter representation of any operator, alternatively an operator $\hat J_{n_i,n_j}$, diagonal in the local particle number operators $\hat n_i$ and containing the density-dependent coefficients, can be constructed.

To transform the total single-particle tunneling Hamiltonian $\mathcal{H}_J=\mathcal{H}_t+\mathcal{H}_{ U,\mbox{\footnotesize nn}}^J$ to the density-dependent parameter form of Eq.~(\ref{EQ:HJ_density_dep}), we insert the unit operator of the low-energy subspace on both the left and the right. For one given operator term, such as $\mathcal A_{i,j}=a_{i,\alpha_1}^{\dag} a_{i,\alpha_2}^{\dag} a_{i,\alpha_3}^{\phantom{\dag}} a_{j,\alpha_4}^{\phantom{\dag}}$, the expectation reduces to a product of expectation values at different sites
\spl{
        {_i}\bra{\psi_0^{(n_i')}}&{_j}\bra{\psi_0^{(n_j')}} A_{i,j} \ket{\psi_0^{(n_i)}}_i \ket{\psi_0^{(n_j)}}_j=\delta_{n_i',n_i+1} \delta_{n_j',n_j-1} \\
        &\times \sqrt{n_j(n_i+1)} n_i \,  f_{(\alpha_3)(\alpha_2,\alpha_1)}^{{(n_i)}^*} \,  f_{\alpha_4}^{(n_j)} ,
}
where the coefficients  $f_{(\alpha_3)(\alpha_2,\alpha_1)}^{{(n_i)}^*}$ are defined in Eq.~(\ref{EQ:def_gerneralized_f}). On all other lattice sites different from $i$ or $j$, $\mathcal A_{i,j}$ acts as the unit operator. The same procedure can be used for $\mathcal{H}_t$. We furthermore use the fact that for a time reversal symmetric system, all Wannier functions can be chosen purely real and the matrix element $U_{\alpha_1,\alpha_2,\alpha_3,\alpha_4}^{(i_1,i_2,i_3,i_4)}$ is invariant under the 24 possible permutations of index pairs $(i_n,\alpha_n)$. Together with the discrete translational symmetry we thus have $U_{\alpha_1,\alpha_2,\alpha_3,\alpha_4}^{(i,j,j,j)}=U_{\alpha_4,\alpha_2,\alpha_3,\alpha_1}^{(i,i,i,j)}$ and upon relabeling the summation indices $\alpha_1 \leftrightarrow \alpha_4$ obtain the total density dependent single-particle tunneling parameter after collecting all terms of $\mathcal{H}_J$

\spl{
&J_{n_i,n_j}=J_{n_i,n_j}^{t}+J_{n_i,n_j}^{U} \\
&=\sum_{\alpha} t^{(\alpha)} {f_{\alpha}^{(n_j)}} {f_{\alpha}^{(n_i+1)}}^* + \sum_{\alpha_1,\alpha_2,\alpha_3,\alpha_4} U_{\alpha_1,\alpha_2,\alpha_3,\alpha_4}^{(i,i,i,j)}\\
&\times \left[ n_i \, f_{(\alpha_3)(\alpha_2,\alpha_1)}^{{(n_i)}^*} f_{\alpha_4}^{(n_j)} +(n_j-1) f_{(\alpha_2)(\alpha_3,\alpha_1)}^{{(n_j-1)}} {f_{\alpha_4}^{(n_i+1)}}^*  \right].
}
The first term contains all multi-orbital contributions from the single-particle lattice Hamiltonian, whereas the second term contains the nearest-neighbor couplings originating directly from the two-body interaction Hamiltonian $\mathcal{H}_{\mbox{\tiny int}}$, which are referred to as \textit{non-linear tunneling} in \cite{Mering11} and \textit{bond-charge tunneling} in \cite{Luehmann11}. The former are plotted as solid lines, whereas the latter are the dotted lines in Fig.~(\ref{FIG:density_dependent_comb}) A and D. For moderate lattice depths $s\lesssim 17 E_r$ both contributions $J_{n_i,n_j}^{t}$ and $J_{n_i,n_j}^{U}$ are negative in sign and favor a condensation in the $k=0$ mode. In contrast, at larger lattice depths, the sign of the contribution from the interaction $J_{n_i,n_j}^{U}$ changes, favoring a condensation in the $k=\frac{\pi}{a}$ mode, competing with the $J_{n_i,n_j}^{t}$ processes. In the regime we considered, the single-particle multi-orbital terms outweigh the interaction terms for reasonably deep lattices. Compared to the bare single-particle tunneling amplitudes, the resulting effective single-particle tunneling is changed on the order of $60\%$ for strong lattices. This effect is enhanced by using Feshbach resonances to adjust the scattering length $a_s$.

\subsection{Two-particle correlated hopping}
\label{SEC:dd_two_particle_tunneling}

For a density-dependent representation of the two-particle tunneling parameter, they are represented by an additional term in the total Hamiltonian
\spl{
\label{EQ:HI_density_dep}
\mathcal{H}_{ U,\mbox{\footnotesize nn}}^I=&\sum_{\langle i,j\rangle} \sum_{n_i,n_j} I_{n_i,n_j} \sqrt{(n_i+1)(n_i+2)}\, \ket{\psi_0^{(n_i+2)}}_i {_i}\bra{\psi_0^{(n_i)}}\\
& \otimes  \sqrt{n_j(n_j-1)}\, \ket{\psi_0^{(n_j-2)}}_j {_j}\bra{\psi_0^{(n_j)}} + \mbox{h.c.}
}
The same procedure as described in Sec.~(\ref{SEC:single_particle_tunneling}) can be used to obtain the density-dependent two-particle tunneling coefficients
\spl{
I_{n_i,n_j}=\sum_{\alpha_1,\alpha_2,\alpha_3,\alpha_4}  U_{\alpha_1,\alpha_2,\alpha_3,\alpha_4}^{(i,i,j,j)} f_{\alpha_1,\alpha_2}^{{(n_i+1)}^*} f_{\alpha_3,\alpha_4}^{(n_j-1)}
}
As shown in Fig.~\ref{FIG:density_dependent_comb} B and E, the two-particle tunneling amplitudes are exponentially sensitive on the lattice depth $s$ and can become very strong on the nearest-neighbor energy scale, set by the single-particle tunneling $J$, even exceeding this by an order of magnitude for the strongly interacting case $g=5 g_{vac}$ that we considered. The density-dependent amplitude $I_{n_i,n_j}$ furthermore increases with increasing occupation numbers and for reasonably strong interactions ($g \gtrsim 1.5 g_{vac}$ for $^{87}$Rb in a 738nm lattice) the dependence on the interaction strength $g$ is approximately linear. 

We point out that a transformation between the density-induced and density-dependent two-particle hopping amplitudes exists, which is of the form of a second order tensor with the  $\mathcal{B}$ matrix defined in Eq.~(\ref{EQ:B_matrix})
\begin{align}
\label{EQ:2P_tun_tensor_trafo}
	I_{n_i,n_j}&=\sum_{m_1,m_2=1}^\infty \mathcal{B}_{n_1+1,m_1} \mathcal{B}_{n_2-1,m_2} R_{m_1,m_2}\\
	R_{m_1,m_2}&=\sum_{n_i=0,n_j=2}^\infty (\mathcal{B}^{-1})_{m_1,n_i+1}(\mathcal{B}^{-1})_{m_2,n_j-1} I_{n_i,n_j}.
\end{align}

\subsection{nearest-neighbor interactions}
\label{SEC:dd_nn_interactions}
The nearest-neighbor interaction Hamiltonian in its density-dependent parameter representation reads
\spl{
\label{EQ:H_nn_int_density_dep}
\mathcal{H}_{ U,\mbox{\footnotesize nn}}^{\mbox{\footnotesize int}}=&\sum_{\langle i,j\rangle} \sum_{n_i,n_j} V_{n_i,n_j}\, n_i\, \ket{\psi_0^{(n_i)}}_i {_i}\bra{\psi_0^{(n_i)}}\\
& \otimes  n_j\, \ket{\psi_0^{(n_j)}}_j {_j}\bra{\psi_0^{(n_j)}} 
}
with the coefficients
\spl{
V_{n_i,n_j}=4\sum_{\alpha_1,\alpha_2,\alpha_3,\alpha_4}  U_{\alpha_1,\alpha_2,\alpha_3,\alpha_4}^{(i,i,j,j)} f_{(\alpha_1)(\alpha_2)}^{{(n_i)}} f_{(\alpha_3)(\alpha_4)}^{{(n_j)}}
}
Note that the coefficient $f_{(\alpha_1,\alpha_2)}^{(n)}$ in this case is exactly the local single-particle density matrix in the multi-orbital Wannier representation for the local many-particle ground state $\ket{\psi_0^{(n)}}$. The amplitudes $V_{n_i,n_j}$ scaled by the bare single-particle tunneling $J$ depend exponentially on the lattice depth $s$, as shown in Fig.~\ref{FIG:density_dependent_comb} C. Note that while $V_{1,1}/J$ decreases with increasing $s$, higher-order terms $V_{n_i,n_j}/J$ with $n_i,n_j>1$ grow exponentially. These also become very strong and can become more relevant than the single-particle hopping elements at large $s$, which is of relevance for processes in the Mott insulating regime.
An analogous second order tensor transformation property as Eq.~(\ref{EQ:2P_tun_tensor_trafo}) applies to the nearest-neighbor interaction amplitudes
\begin{align}
\label{EQ:nn_int_tensor_trafo}
	V_{n_i,n_j}&=\sum_{m_1,m_2=1}^\infty \mathcal{B}_{n_1,m_1} \mathcal{B}_{n_2,m_2} W_{m_1,m_2}\\
	W_{m_1,m_2}&=\sum_{n_i,n_j=1}^\infty (\mathcal{B}^{-1})_{m_1,n_i}(\mathcal{B}^{-1})_{m_2,n_j} V_{n_i,n_j}.
\end{align}

\subsection{On-site energies}
\begin{figure}[b]
\includegraphics[width=\linewidth]{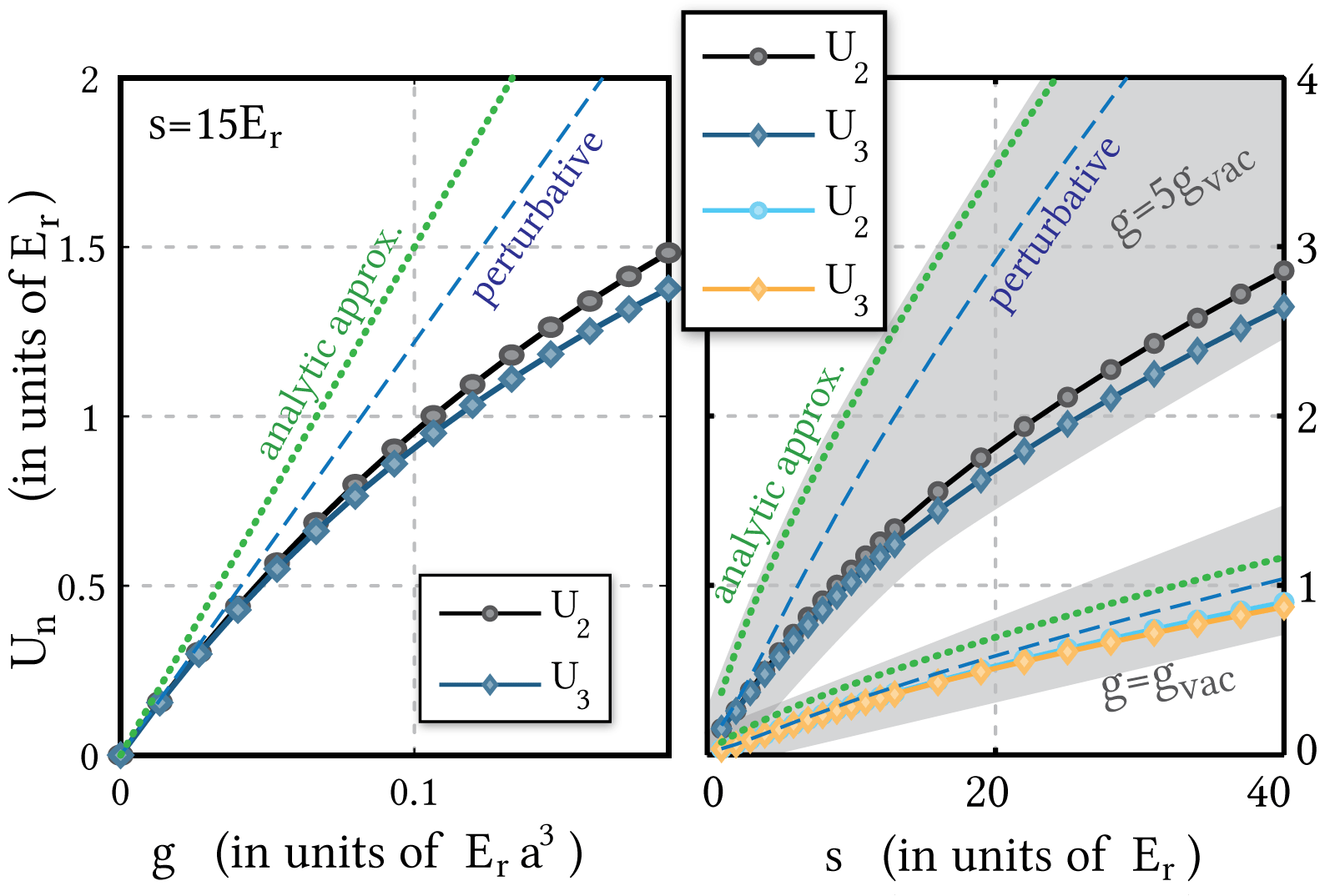}
        \caption
        {\label{FIG:density_dependent_local}
(Color online) The density-dependent local parameters as a function of the interaction strength $g$ at constant $s=15E_r$ (left) and as a function of the lattice depth $s$ at two fixed values of the interaction strength $g=g_{\mbox{\tiny vac}}$ and $g=5g_{\mbox{\tiny vac}}$. Shown as green dotted lines is the analytic approximation $U=\sqrt{8\pi} \frac{a_s}{a} \left({s}/{E_r}\right)^{3/4} E_r$. The perturbative result from a single-particle band structure calculation is shown as blue dashed lines.}
\end{figure}
The exact diagonalization procedure yields the Hamiltonian containing all on-site terms in the form
\spl{
\PlowE \mathcal{H}_{\mbox{\footnotesize loc}}  \PlowE & = \sum_i \sum_n E_0^{(n)}  \ket{\psi_0^{(n)}}_i {_i}\bra{\psi_0^{(n)}} 
}
with the $n$-particle ground state energies $E_0^{(n)}$ being the numerically found lowest eigenvalues. Within a Bose-Hubbard formulation of the same Hamiltonian with density-dependent interaction parameters $U_n$, each on-site term of this Hamiltonian is to be expressed in the form
\spl{
  \PlowE  \mathcal{H}_{\mbox{\footnotesize loc}}^{(i)}  \PlowE & = \sum_n  (\epsilon^{(0)} - \mu) n \ket{\psi_0^{(n)}}_i {_i}\bra{\psi_0^{(n)}} \\
  &+\sum_n  \frac{U_n}{2} n(n-1) \ket{\psi_0^{(n)}}_i {_i}\bra{\psi_0^{(n)}} 
}
and is of course identical to the single-particle energy formally obtained from the many-particle diagonalization
\begin{equation}
	E_0^{(1)}=\epsilon^{(0)} - \mu.
\end{equation}
Since for $n=1$ interactions do not play a role and the particle is in the lowest Wannier orbital only, thus only the lowest band single-particle energy $\epsilon^{(0)}$ contributes. By subtracting the single-particle energy shift for higher occupations $n>1$, the density-dependent interaction parameter is found to be
\begin{equation}
	U_n=2\frac{E_0^{(n)}-(\epsilon^{(0)} - \mu)n}{n(n-1)}.
\end{equation}
In the limit of very weak interactions (compared to the single-particle hopping energy), where the interaction can be treated perturbatively, the parameters $U_n$ become independent of the local density $n$ and all coincide with the usual interaction energy $U$, as shown in Fig.~\ref{FIG:density_dependent_local}.

\section{correlations vs. orbital deformation}
\label{SEC:higher_order_correlations}

\begin{figure}[b]
\includegraphics[width=\linewidth]{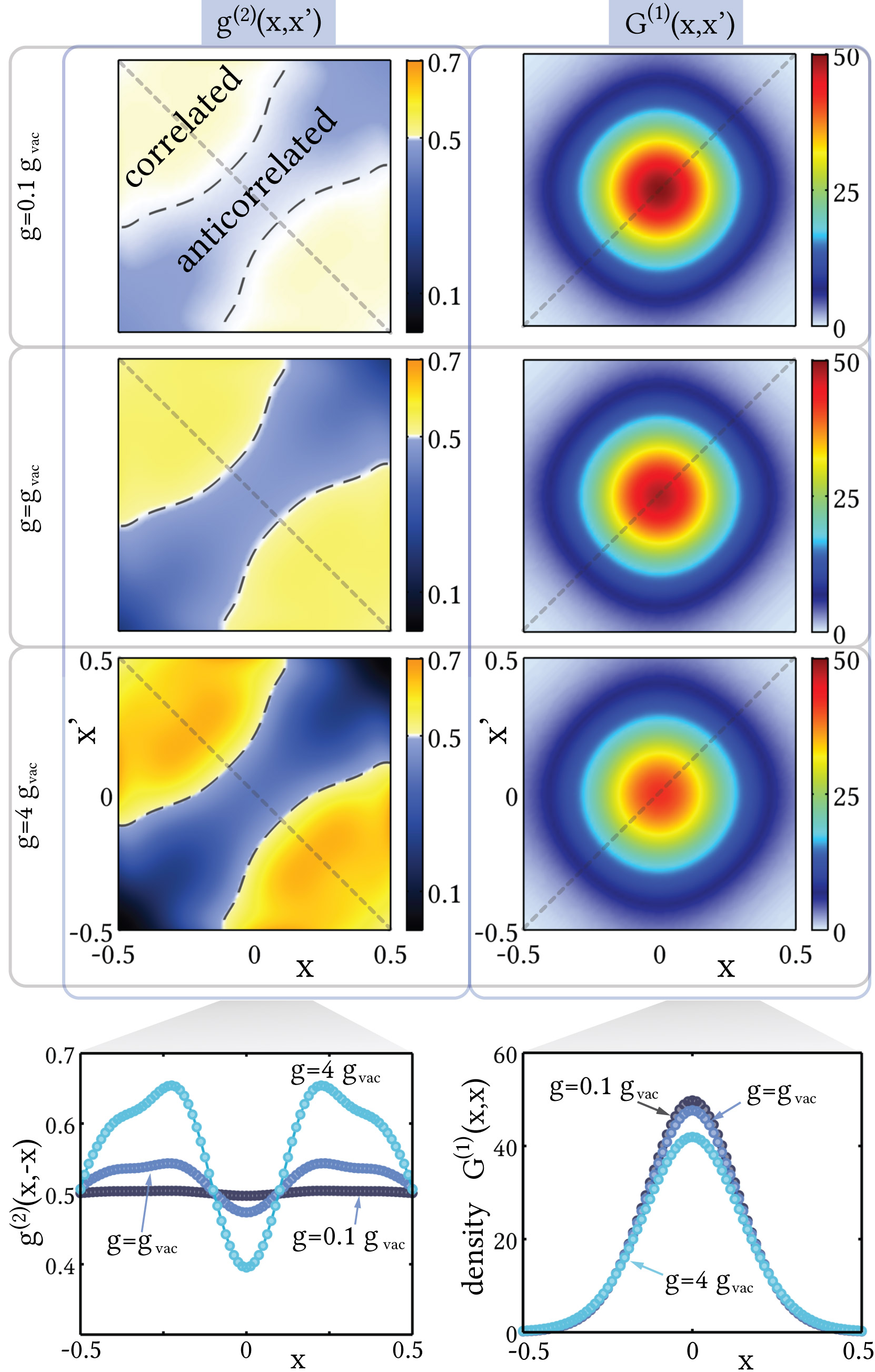}
            \caption
            {\label{fig:corr_fcts}
(Color online) Correlation functions $g^{(2)}(x,x')$ (left column) and single-particle density matrix $G^{(1)}(x,x')=\ev{\Psi^\dag(x) \Psi(x')}$ (right column) in real-space at $y=y'=z=z'=0$ for a single site in units of the lattice constant $a$. Results are shown for $n=2$  $^{87}\mbox{Rb}$ atoms in a $768$nm $s=10E_r$ lattice for three different interaction strengths, up to four times the vacuum interaction strength $g_{\mbox{\tiny vac}}$. The lower graphs are cuts along the dotted lines in the plots above. With increasing interaction strength, the atoms mutually avoid each other, as can be seen in the decrease of $g^{(2)}(x,x)$ along the diagonal ($x=x'$) and an increase for $|x-x'|\gtrapprox 0.2$, compared to the noninteracting case where $g^{(2)}(x,x')=0.5$ for $n=2$. The effect on the density distribution along one direction is significantly weaker, with the main effect being a reduction of density at the center of the lattice site.
}\end{figure}

We demonstrate now that the deformation of single-particle orbitals by the interactions is not the main effect to lower the local on-site energy. The higher-order correlations, which cannot be understood as such a deformation, are the dominant effect to lower the energy. Therefore, a single-particle picture and wave functions are not sufficient for understanding the effect of interactions on the local level, since entanglement becomes important. This can best be seen in the two-particle correlation function in Fig.~(\ref{fig:corr_fcts}).

Given the state $\ket{\psi_0^{(n)}}_i$, the local single-particle density matrix is Hermitian and can thus be expressed in terms of its orthogonal eigenvectors (corresponding to single-particle states in the respective basis) $\phi_{\alpha}^{(l)}$ and the corresponding real, positive eigenvalues $\lambda_l$
\spl{
\rho_{\alpha,\alpha'}&=\bra{\psi_0^{(n)}}  a_{i,\alpha}^{\dag} a_{i,\alpha'}^{\phantom{\dag}}   \ket{\psi_0^{(n)}}\\
&=\sum_l \lambda_l \, \phi_{\alpha}^{(l)} \, {\phi_{\alpha'}^{(l)}}^*.
}
The eigenvalues $\lambda_l$ and the associated single-particle states do not, of course, depend on the basis in which the single-particle density matrix is evaluated.

We can now construct an artificial state
\begin{equation}
\label{EQ:constr_uncorr_mp_state}
\ket{\psi_{uc}}=\sum_l \sqrt{\frac{\lambda_l}{n}}  \, \frac{1}{\sqrt{n!}} (d_l^\dag)^n \, \ket 0
\end{equation}
for comparison, which leads to the identical single-particle density matrix, but does not contain the higher-order correlations.
\begin{figure}[t]
\includegraphics[width=0.85\linewidth]{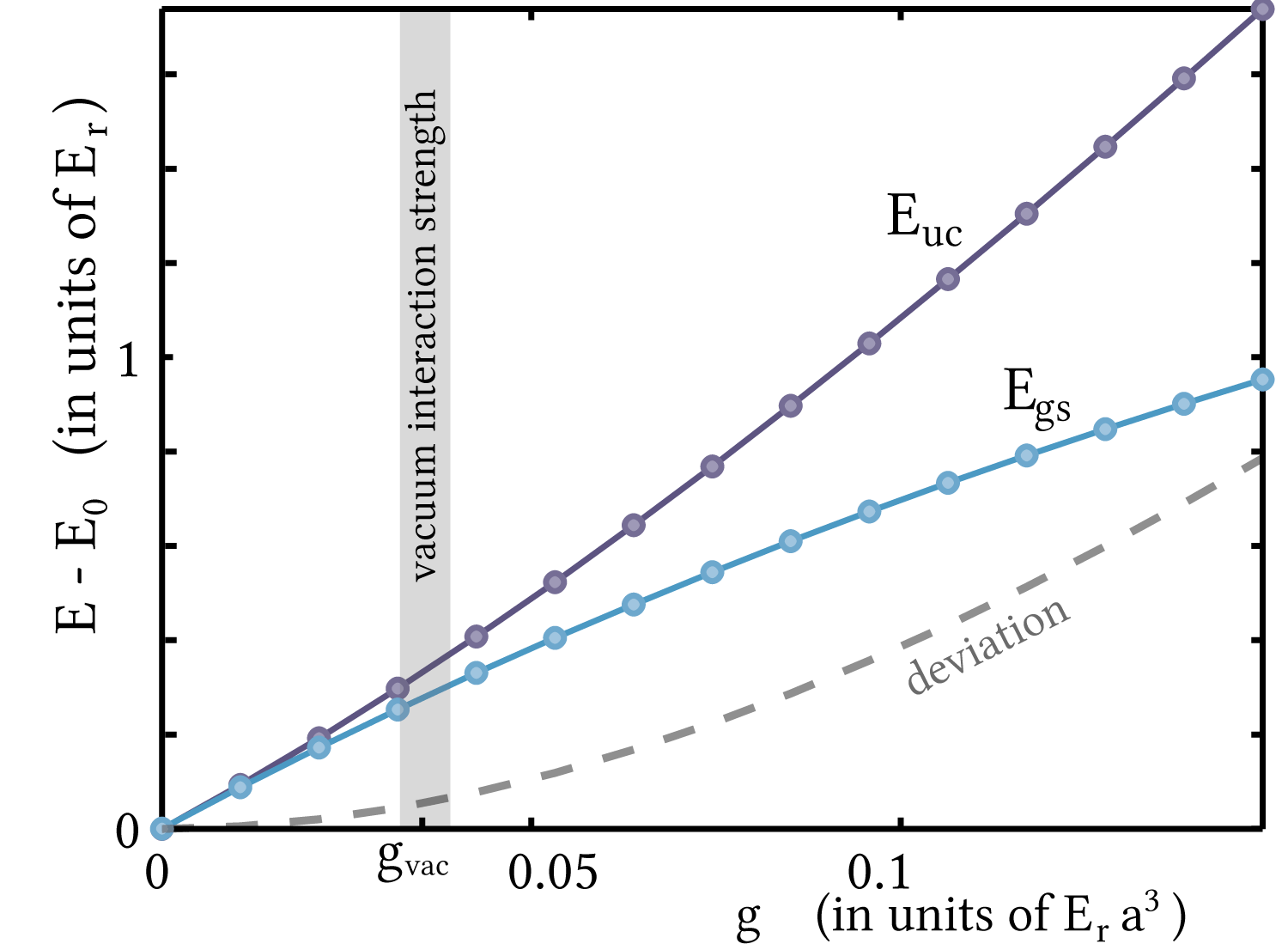}
            \caption
            {\label{Fig:uncorr_E_comparison}
(Color online) Comparison of the true many-body local ground state energy $E_{gs}$ and the energy $E_{uc}$ of the artificially created state $\ket{\psi_{uc}}$ of Eq.~(\ref{EQ:constr_uncorr_mp_state}) with the same single-particle density matrix (i.e. the same broadened orbitals and their occupation), but no higher-order correlations. The deviation between the two becomes very significant with increasing interaction strength and is well in the experimentally observable regime (here for $s=10$, but the effect becomes stronger with increasing lattice depth), showing that the usual simple picture of broadened single-particle orbitals is insufficient to explain the effects of interactions on the local many-body state.
}\end{figure}

Here we defined the creation operators for the eigenstates of the single-particle density matrix $d_l^\dag = \sum_\alpha  \phi_{\alpha}^{(l)}  a_{i,\alpha}^{\dag}.$ The state in Eq.~(\ref{EQ:constr_uncorr_mp_state}) can be thought of as having the identical single-particle properties as the true local ground state, and would be the most natural many-particle state for thinking in terms of spatially broadened single-particle orbitals due to the interactions, as commonly referred to in literature \cite{Will10}. It does not, however, contain the same higher-order correlations as the original state, for instance the two-particle correlation function $G^{(2)}_{l,l'}=\bra{\psi_{uc}} d_{l}^\dag \, d_{l'}^\dag  d_{l'}^{\phantom{\dag}} d_{l}^{\phantom{\dag}} \ket{\psi_{uc}}=\lambda_l \, \delta_{l,l'}$ for $n=2$ particles or more. In Fig.~(\ref{Fig:uncorr_E_comparison}) the energy expectation value of the uncorrelated state $E_{uc}=\bra{\psi_{uc}} \mathcal{H} \ket{\psi_{uc}} $ is compared to the true ground state energy as a function of the interaction strength $g$. Since the above constructed state $\ket{\psi_{uc}}$ becomes the true ground state in the noninteracting limit, the energies agree in this limit. However, significant deviations arise at finite interaction strengths $g$, indicating that the simple picture of spatially broadened single-particle orbitals cannot explain the main energy reduction mechanism.

The significant change of the on-site many-body state lies in the higher-order correlation functions, with particles mutually reducing their spatial overlap. It is not contained in and cannot be understood on the single-particle level (since all single-particle properties of the local state are contained in $\rho_{\alpha,\alpha’}^{(i)}$). To substantiate this point, we calculated the normalized second-order correlation function
\spl{
g^{(2)}(\br,\br')=\frac{\bra{\psi} \Psi^\dag(\br)\, \Psi^\dag(\br') \, \Psi(\br') \, \Psi(\br)  \ket{\psi}  } {( \bra{\psi} \Psi^\dag(\br)\,  \Psi(\br)  \ket{\psi}  )(\bra{\psi} \Psi^\dag(\br')\,  \Psi(\br')  \ket{\psi}) }
}
for a local interacting two-particle ground state $\ket{\psi_0^{(2)}}$, shown in Fig.~(\ref{fig:corr_fcts}). The normalized second order correlation function can be understood as a conditional probability: for the noninteracting state $\ket{n=2}$ we have $g^{(2)}(\br,\br')=1-\frac 1 n=\frac 1 2$, which is our reference and which we refer to as uncorrelated by interactions. A value of $g^{(2)}(\br,\br')<\frac 1 2$ indicates a reduced probability for a particle to be found at location $\br'$ if another particle is located at $\br$ or vice versa and is therefore anticorrelated in this sense. This anticorrelation can be seen along the diagonal line $x=x'$ in Fig.~(\ref{fig:corr_fcts}), where two repulsively interacting atoms have a reduced probability of being found at the same spatial position $x=x'$. Since all particles are restricted to occupy Wannier orbitals at the same site, the conditional probability $g^{(2)}(\br,\br')$ has to be increased elsewhere, i.e., correlated. With increasing interactions, the specific shape of $g^{(2)}(\br,\br')$ is independent of the interaction strength $g$ in this regime in the sense that the deviation from the uncorrelated case scales linearly with $g$. This can clearly be seen by comparing the different functions in the left column of Fig.~(\ref{fig:corr_fcts}). In contrast, the density profile shown in the right column is only slightly changed by the interactions. The main effect is a reduction of the maximal density at the center of the lattice site, whereas only a minimal broadening of the density profile is visible.

\section{Conclusions}
We have formulated a systematic derivation of an effective low-energy, single-band basis for ultracold bosonic atoms in optical lattices in the presence of interactions. Some properties intrinsic to our formalism, such as density-dependent interaction parameters or the appearance of effective multibody interactions have been previously discussed and experimentally confirmed. We introduce ladder operators fulfilling bosonic commutation relations within the new dressed-band basis, which are shown to be the bosonic operators used within an effective Bose-Hubbard model for the system. It is however shown that these are not the original lowest band Wannier creation and annihilation operators beyond lowest the order and we derive a simple prescription for the transformation of arbitrary operators into the new low-energy dressed-band basis. These transformations are used to systematically treat all terms in the interacting lattice Hamiltonian and give rise to multibody-induced single and pair particle tunneling, as well as multibody local and nearest-neighbor interactions. The amplitudes for these processes are calculated and compared to renormalized parameters in the density-dependent representation of the Bose-Hubbard model. The latter formulation, although fully equivalent, is, however, less favorable for the treatment with a number of common theoretical methods, since it contains an external summation over the set of all $n$-particle states at each lattice site. We furthermore show that the commonly used single-particle picture of spatially broadened Wannier orbitals cannot describe the observed energy reduction of the local many-body state. The relevant mechanism is mutual avoidance of the various atoms at a given lattice site, which is a many-particle effect contained only in the higher-order correlation functions.

\begin{acknowledgments}
We thank M. Buchhold, D. Cocks, A. J. Daley, M. Fleischhauer, O. J\"urgensen, D.-S. L\"uhmann, and S. Will for useful discussions.  This work was supported by the DFG via Forschergruppe FOR 801 and Sonderforschungsbereich SFB/TR 49. W.H. acknowledges the hospitality of the Aspen Center of Physics during the final stage of this work, supported by the National Science Foundation under Grant No. 1066293.
\end{acknowledgments}

\begin{appendix}
\section{Parity Symmetry}
\label{SEC:appendix_parity}

Here we prove that the local interacting Hamiltonian preserves the multiparticle parity along each direction at a given site $i$ and drop the site index for this section.
The local Wannier orbitals are either fully spatially symmetric or antisymmetric in all of the three spatial dimensions, thus rendering the multiparticle parity operator diagonal in the Wannier Fock space representation. Since the local lattice Hamiltonian is also diagonal in this representation, we can directly infer $[\mathcal{H}_{\epsilon}, Q^{(x)}]=0$.

We now focus on proving the second relation $[\mathcal{H}_{ U,\mbox{\footnotesize loc}}, Q^{(x)}]=0$. This is equivalent to the statement that both operators share a common basis of eigenvectors, or, equivalently, that in a given basis one operator is block diagonal, with non-diagonal blocks only within those subspaces where the other operator is a scalar multiple of the identity. We take the basis of Wannier Fock states and consider the expression
\begin{equation}
          (\mathcal{H}_{ U,\mbox{\footnotesize loc}} Q^{(x)}- Q^{(x)} \mathcal{H}_{ U,\mbox{\footnotesize loc}}) \ket{n_{\alpha_1},\ldots,n_{\alpha_M}}.
\end{equation}
Since the state is an eigenstate of $ Q^{(x)}$, the first term corresponds to $\lambda \mathcal{H}_{ U,\mbox{\footnotesize loc}}\ket{n_{\alpha_1},\ldots,n_{\alpha_M}}$, where
\begin{equation}
          \lambda = (-1)^{\sum_{\alpha_x=1,3,\ldots} \sum_{\alpha_y,\alpha_z} n_{\alpha_x,\alpha_y,\alpha_z}}
\end{equation}
is the eigenvalue belonging specifically to this state.

We now consider the second term
\begin{equation}
\label{eq:parity_second_term}
          Q^{(x)} \frac g 2
          \sum_{\alpha_1,\alpha_2,\alpha_3,\alpha_4} U_{\alpha_1,\alpha_2,\alpha_3,\alpha_4}   a_{\alpha_1}^\dag a_{\alpha_2}^\dag a_{\alpha_3}^{\phantom{\dag}} a_{\alpha_4}^{\phantom{\dag}}             
          \ket{n_{\alpha_1},\ldots,n_{\alpha_M}}
\end{equation}
and use the property that the interaction matrix element $U_{\alpha_1,\alpha_2,\alpha_3,\alpha_4}=\prod_{d=x,y,z} U_{\alpha_{1,d},\alpha_{2,d},\alpha_{3,d},\alpha_{4,d}}$ factorizes into a product of terms from the individual dimensions. One such term,
\begin{equation}
U_{\alpha_{1,x},\alpha_{2,x},\alpha_{3,x},\alpha_{4,x}}=\int dx \, w_{\alpha_{1,x}}^* (x)\, w_{\alpha_{2,x}}^*(x) \, w_{\alpha_{3,x}}(x)\, w_{\alpha_{4,x}}(x)
\end{equation}
vanishes if it contains an odd number of odd functions. Consequently, all non-vanishing states created in the sum in Eq.~(\ref{eq:parity_second_term}) are also eigenstates of $ Q^{(x)}$ to the same eigenvalue $\lambda$ as the initial state.

As the Bloch Fock states constitute a complete basis, Eq.~(\ref{eq:parity_second_term}) holds on an operator level, i.e., $[\mathcal{H}_{ U,\mbox{\footnotesize loc}} , Q^{(x)}]=0$.

\end{appendix}

\end{document}